\documentclass[preprintnumbers, floatfix, preprintnumbers, letterpaper, superscriptaddress,nofootinbib, onecolumn]{revtex4}
\usepackage{array}
\usepackage[colorinlistoftodos]{todonotes}

\newcolumntype{P}[1]{>{\centering\arraybackslash}p{#1}}
\newcolumntype{M}[1]{>{\centering\arraybackslash}m{#1}}
\usepackage{graphicx}
\usepackage{microtype}
\usepackage{amsmath}
\usepackage{amssymb}
\usepackage{subfigure}
\usepackage{hyperref}
\usepackage{url}
\usepackage{xcolor}
\usepackage{color}
\usepackage{mathrsfs}
\usepackage{calrsfs}
\usepackage{amsfonts}
\usepackage{latexsym}
\usepackage{ragged2e}
\usepackage{epsfig}
\usepackage{textcomp}
\usepackage{eufrak}
\usepackage{tabularx}
\usepackage{eucal}
\usepackage{epsfig}
\usepackage{amsmath} 
\usepackage[utf8]{inputenc}
\usepackage{fourier} 
\usepackage{array}
\usepackage{makecell}
\usepackage{longtable}

\definecolor{vividviolet}{rgb}{0.62, 0.0, 1.0}
\definecolor{amaranth}{rgb}{0.9, 0.17, 0.31}
\definecolor{palatinateblue}{rgb}{0.15, 0.23, 0.89}
\definecolor{brightpink}{rgb}{1.0, 0.0, 0.5}
\definecolor{cornflowerblue}{rgb}{0.39, 0.58, 0.93}
\definecolor{deepcarminepink}{rgb}{0.94, 0.19, 0.22}
\definecolor{radicalred}{rgb}{1.0, 0.21, 0.37}

\hypersetup{ linktoc=all,
	colorlinks, linkcolor={palatinateblue},
	citecolor={brightpink}, urlcolor={amaranth}
}

\graphicspath{{Images/}}

\def\sideremark#1{\ifvmode\leavevmode\fi\vadjust{\vbox to0pt{\vss
			\hbox to 0pt{\hskip\hsize\hskip1em
				\vbox{\hsize1.3cm\tiny\raggedright\pretolerance10000
					\noindent #1\hfill}\hss}\vbox to8pt{\vfil}\vss}}}%
\def\beq{\begin{equation}}
\def\eeq{\end{equation}}

\begin{document}

\tolerance=5000

\title{Assessing  the foundation  and applicability of some dark energy fluid models in the Dirac-Born-Infeld framework}

\author{Muhsin Aljaf}
\email{mohsen@mail.ustc.edu.cn}
\affiliation{Department of Physics, Oakland University, Rochester, MI 48309, USA}
\affiliation{Department of Physics, College of Education, University of Garmian  Kurdistan region, Iraq}

\author{Daniele Gregoris}
\email{danielegregoris@libero.it}
\affiliation{School of Science, Jiangsu University of Science and Technology, Zhenjiang 212003, China}

\author{Martiros Khurshudyan}
\email{ khurshudyan@ice.csic.es, martiros.khurshudyan@csic.es }
\affiliation{
 Consejo Superior de Investigaciones Cient\'{\i}ficas, ICE/CSIC-IEEC,
Campus UAB, Carrer de Can Magrans s/n, 08193 Bellaterra (Barcelona) Spain }

\begin{abstract}

In this paper, we will deepen the understanding of some fluid models proposed by other authors for the description of dark energy. Specifically, we will show that the so-called (Modified) Berthelot fluid is the hydrodynamic realization of the free Dirac-Born-Infeld theory and that the Dieterici fluid admits a non-relativistic $k$-essence formulation; for the former model the evolution of the scalar field will be written in terms of some cosmographic parameters. The latter model will also be tested using Machine Learning algorithms with respect to cosmic chronometers data, and results about the dynamics at a background level will be compared with those arising when other fluids (Generalized Chaplygin Gas and Anton-Schmidt) are considered. Due to some cosmic opacity effects, the background cosmology of universes filled by these inequivalent fluids, as they arise in physically different theories, may not be enough for discriminating among them. Thus, a perturbation analysis in the long-wavelength limit is carried out  revealing a rich variety of possible behaviors. It  will  also be shown  that the free Dirac-Born-Infeld theory cannot account for flat galactic rotation curves, and therefore we derive an appropriate relationship between the scalar field potential and the brane tension for achieving this goal; this provides an estimate for the dark matter adiabatic speed of sound inside the halo consistent with other literature. A certain relationship between the Newtonian gravitational potential within the galaxy  and the Lagrangian potential in the non-relativistic regime for the (Modified) Berthelot fluid will also be enlightened.

\end{abstract}


\maketitle

\section{Introduction}\label{sec:INT}

The existence of a dark energy fluid has been claimed for accounting for the accelerated expansion of the Universe  \cite{sup1,sup2}. Its existence is compatible also with the theoretical description of the cosmic microwave background \cite{cmb1,cmb2}, the baryon acoustic oscillation (BAO) spectra \cite{bao1,bao2}, and the cosmic evolution \cite{HZ2,HZ3}. The minimal proposal for the modelling of dark energy, the cosmological constant,  is mathematically consistent with the symmetries of general relativity, but this approach would violate the causality principle, should we interpret the cosmological constant as a fluid with $p=-\rho$ \cite{Ref1}. 
On the other hand, astrophysical observations suggest that our Universe should also be filled with dark matter, which can be macroscopically pictured as pressureless dust, regardless of its elementary particle foundation in terms of either massive neutrinos, sterile neutrinos, axions, axinos, gravitinos, or neutralinos \cite{Ref2,Ref3,Ref4,Ref5,Ref6,Ref7,Ref8} (see, e.g., \cite{annrev} for a review of different candidates of dark matter particles). Some amount of non-baryonic matter is needed to account for the rotation curves of galaxies and to address the problem of the formation of astrophysical structures. In fact,  a local collapse of cosmic material requires dark matter to dominate over dark energy in the past cosmological epochs, as it can be seen in the behavior of the spectrum of mass fluctuations at short wavelengths (see \cite{bertone} and references therein).

The Generalized Chaplygin Gas is a very successful proposal for a unified description of these two cosmic fluids, and it comes with the good features of restoring causality and predicting a transition from an epoch dominated by dark matter to a subsequent one dominated by dark energy. This permits us to account for the existence of astrophysical structures inside an accelerated expanding  Universe, and the sturdiness of this paradigm has been confirmed by a large number of model testing investigations, as summarized in \cite{chapI, Sergei}. 
Furthermore, it has been discovered that the Generalized Chaplygin Gas is not only an ad-hoc phenomenological framework because it  arises also from the Nambu-Goto theory with soft corrections (i.e. logarithmic in the kinetic energy of the scalar field) for the dynamics of a 2-dimensional brane \cite{gene1}, and the properties of its symmetry group have been widely discussed \cite{jackiw}. Nevertheless, also the Dieterici fluid can realize a dark matter-dark energy phase transition \cite{dieterici1,dieterici2}, whose quantitative predictions with respect to cosmic chronometers data will be assessed in this paper. The reason is that a series expansion in the energy density of the Dieterici equation of state (EoS) may be constituted by Generalized-Chaplygin-Gas-like addenda, making the analysis whether the Nature favours exactly one of those terms, like in a Generalized Chaplygin Gas, or their summation, of crucial importance.
Here we will show that the redshift evolution of the Hubble rate does not necessarily call for an extension of the Generalized Chaplygin Gas into a Dieterici fluid; the data points describing the cosmic history were obtained by applying the differential age method, and the reader can find them reported in \cite{zhang2016}. This technique allows estimating the time derivative of the redshift through the measurements of the different ages of passively-evolving galaxies at different redshifts, which is then used to determine the values of the cosmological parameters, like the Hubble constant and the EoS of the cosmic fluid \cite{jimenez}.  In our analysis, we will forecast the model-free parameters by applying recently developed Machine Learning (ML) algorithms in which we {\it train} a computer system to make predictions by recognizing the properties of the data themselves.
In fact,  our ML approach has already been proven to be a valuable technique for testing the dataset's consistency, searching for new physics and tensions in the data, and placing tighter constraints on the theoretical parameters \cite{revmod}. For example,  they have been employed for taming systematics in the estimates of the masses of clusters, for discriminating between different gravity theories when looking at statistically similar weak lensing maps, in N-body simulations, cosmological parameters inference, dark energy model comparison, supernova classification,  strong lensing probes, and it is expected that they will be useful for the next generation of CMB experiments, (see \cite{revmod} for a report about the state of the art applications of ML approaches over many different branches of Physics). 

Another fluid model which has already been tested as a dark energy candidate is the so-called (Modified) Berthelot fluid \cite{capo}. While being originally proposed for a realistic description of the fugacity properties of hydrocarbons \cite{reos2}, it has received novel attention in recent cosmological applications for example in interacting dark matter - dark energy scenarios in which, in particular, the avoidance of specific cosmological singularities has been assessed \cite{epjc2020}, in black hole cosmic accretion phenomena in the McVittie framework \cite{mcvittie}, and in early-late time unification in quadratic gravity \cite{saikat}. In this paper, we will provide a transparent foundation of this hydrodynamical model by reconstructing its relationship to the Dirac-Born-Infeld theory. We will as well reconstruct the $k$-essence formulation of the aforementioned Dieterici fluid in the non-relativistic regime, which would then be compared with those of the Generalized Chaplygin Gas and of the Anton-Schmidt fluid, which instead belong to the family of the Generalized Born-Infeld theories \cite{poin,avelino}.   We refer also to  \cite{cheva} for the construction of the k-essence formulation for barotropic, logotropic and Bose-Einstein condensates fluid models. Exploiting the relationship between (Modified) Berthelot fluid and free Dirac-Born-Infeld theory we will show that this paradigm is challenged by astrophysical requirements of flat galactic rotation curves: this will call us to derive an appropriate relationship between  scalar field potential and brane tension.

Unfortunately, investigating the dynamics at a background level for these fluids is not enough for concluding which one is the physical theory behind dark energy. A further source of uncertainty comes from possible scattering effects between the photons constituting the light rays,  whose detection constitutes a primary way of testing a certain cosmological model,  and the interstellar medium. This phenomenon is accounted for by introducing a cosmic opacity function that acts as a scaling for the luminosity function. A plethora of different redshift parametrizations for the cosmic opacity has been constructed and investigated \cite{Lee}.  While a non perfectly transparent universe can be a solution to some of the current observational tensions, as small and large scale estimates of the Hubble constant \cite{PyMC3_p2}, this would somehow hide the properties of dark energy, and models based on different fluid pictures may become statistically indistinguishable at the background level. Therefore, for breaking this degeneracy, we will perform a mathematical analysis of the evolution of perturbations in the long-wavelength limit  for epochs in which dark energy dominates over dark matter assuming small deviations from the ideal fluid behavior in which pressure and energy density would be proportional to each other.  We will find a rich variety of behaviors for the different fluid models either exhibiting time increasing or decreasing density contrasts and with or without oscillations; we will be able to provide as well some analytical solutions in a closed form which can serve as a valuable tool for comparison with similar analysis concerning further dark energy models that future literature may come up with.

Our paper is organized as follows: In Sect.\ref{sec:Mod}, we introduce the dynamical equations at a background level for the cosmological models we are interested in, deriving a novel non-relativistic $k$-essence formulation for the Dieterici fluid and proposing a connection between the (Modified) Berthelot fluid and the Dirac-Born-Infeld paradigm for deepening the understanding of the meaning of the free parameters beyond the thermodynamical fluid approach. In Sect.\ref{testing},  we will first review  the foundation of the specific Bayesian Machine Learning (BML) method  we will apply explaining its suitability for our analysis,  and then we will forecast the free parameters of the model based on the Dieterici fluid also validating our results by comparing with observational data. Some degeneracy between physically different (e.g., arising in different frameworks as the Dirac-Born-Infeld vs. the Generalized Born Infeld) descriptions of dark energy will be mentioned in light of cosmic opacity effects in Sect.\ref{opacity}. This will call for the study of the evolution of density perturbations beyond the background level, whose results in the long-wavelength limit  will be exhibited in Sect.\ref{perturbation} for different modelings of the cosmic fluid.  In Sect.\ref{rotation} we will assess the applicability of the models  with respect to flat galactic rotation curves first pointing out and then resolving some shortcomings of the free Dirac-Born-Infeld model.
Finally,   we will comment on the cosmological meaning of our results and conclude in Sect.\ref{sec:conc}.

\section{Background field equations}\label{sec:Mod}

The dynamics at the background level of the Hubble function $H$, and of the energy densities of cold dark matter and dark energy in a flat homogeneous and isotropic universe\footnote{For a recent assessment on the curvature properties of the Universe we refer to \cite{piatto}.} is governed by the evolution equations \cite{peebles}:
\begin{eqnarray}
\label{eq:F1}
 H^{2} \,  = \, \frac{\rho + \rho_{dm}}{3}  \,, \qquad
\dot \rho \,  = \, -3H(\rho+p)  \,, \qquad
 \dot \rho_{dm} \,  = \, -3H\rho_{dm}  \,,
\end{eqnarray}
where  an over dot stands for a derivative with respect to the cosmic time.
For avoiding loss of generality and accounting for the observational properties of the matter power spectrum \cite{zalda}, we will assume that the dark matter abundance is quantified either by $\rho_{dm}$, and by some unified form of dark matter - dark energy in which pressure and energy density are related to each other according to equations of state of the form $p =w(\rho)\rho$. In particular, we want here to establish the field theory foundation of two fluid models previously proposed by some other authors just at the hydrodynamical level and known under the name of
 Dieterici model \cite{dieterici1,dieterici2}
\begin{equation}\label{eq:Pde}
p = A \rho e^{B \rho} \,,
\end{equation}
and of (Modified) Berthelot fluid \cite{capo,reos2}
\beq\label{EOS2}
p=\frac{B \rho}{1+A \rho} \,,
\eeq
with $A$ and $B$ being free parameters.

\subsection{Field theory formulation of the Dieterici fluid}

\begin{figure}[h!]
	\begin{center}$
		\begin{array}{cccc}
		\includegraphics[width=75 mm]{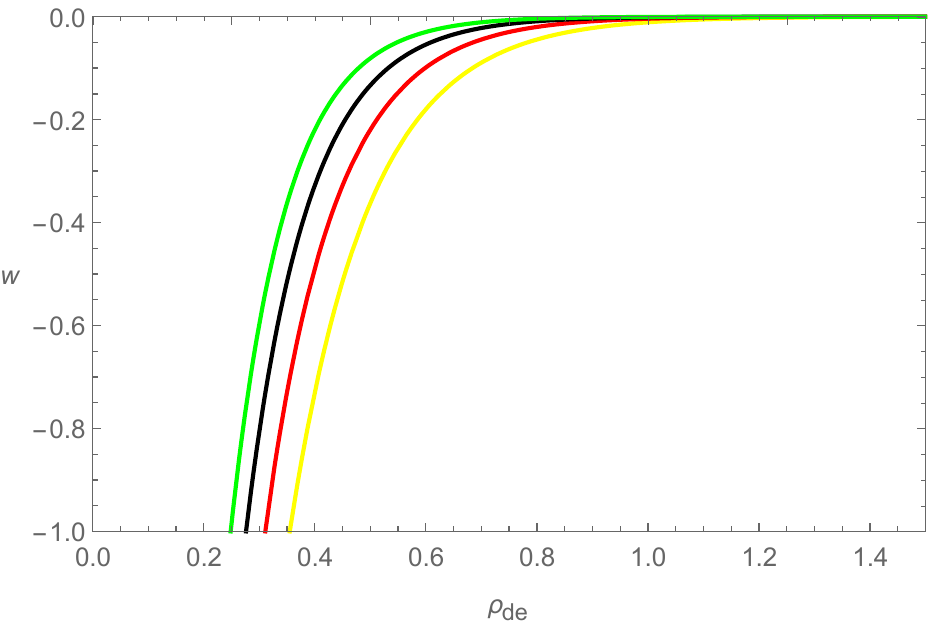}&&
		\includegraphics[width=75 mm]{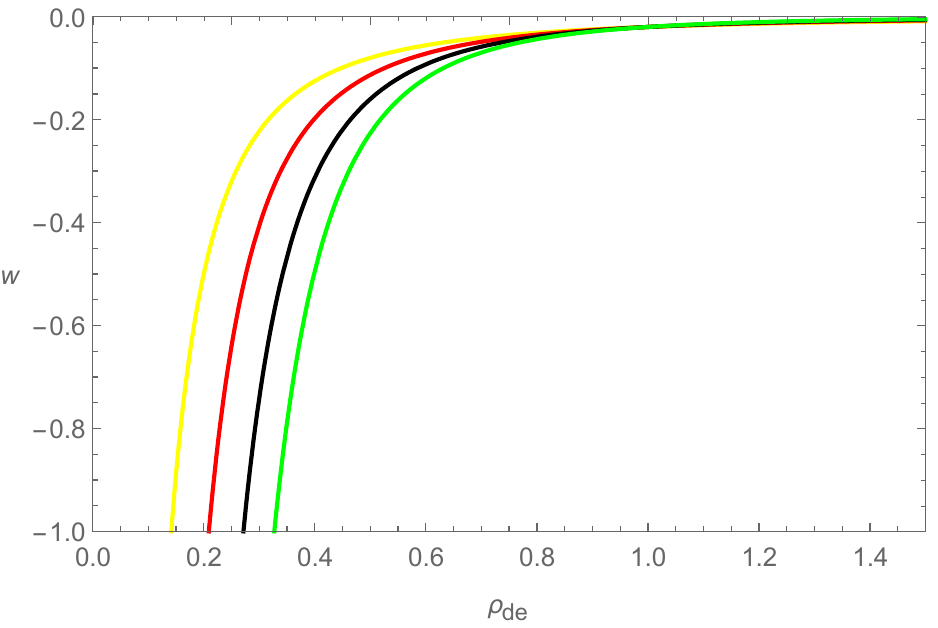}\\
		\end{array}$
	\end{center}
	\caption{The figure displays the comparison between the effective EoS parameter $w={p}/\rho$ of the Dieterici fluid (left panel) and of the  Generalized Chaplygin Gas (right panel).  In the former case  we have fixed $A=-12.0$ and $B=-7.0$ (yellow curve), $B=-8.0$ (red curve), $B=-9.0$ (black curve), and $B=-10.0$ (green curve), while in the latter we have fixed $A=0.02$ and $B=1.0$ (yellow curve), $B=1.5$ (red curve), $B=2.0$ (black curve), and $B=2.5$ (green curve). Both  fluid models interpolate between pressureless dust at high energy density, and a form of dark energy in the opposite regime, and their qualitative similarity is noteworthy.  }
	\label{figeos}
\end{figure}

The Dieterici equation of state (\ref{eq:Pde}) reduces to the Van der Waals model for imperfect fluid at low energy density \cite{dieterici3,dieterici4,dieterici5}. The physical idea at the core of this model is to incorporate the attractive and repulsive contributions to the pressure as
\beq
P=P_{\rm repulsive}e^{-P_{\rm attractive}}\,,
\eeq
rather than
\beq
P=P_{\rm repulsive}+P_{\rm attractive}\,,
\eeq
for getting more realistic estimates of the value of the critical compressibility factor \cite{dietericiaip}.
We are interested in this particular fluid model because it can interpolate between pressureless dust at high densities and a form of dark energy in the opposite regime, as it can be appreciated by looking at its effective equation of state parameter
\beq\label{w}
w:=\frac{p}{\rho}= A e^{B \rho}\,,
\eeq
if we choose negative values for the free parameters $A$ and $B$ \cite{dieterici1,dieterici2}.
We exhibit in Fig.~\ref{figeos} a comparison between the effective EoS parameter $w$ for the Dieterici fluid and the Generalized Chaplygin Gas  (the latter fluid is characterized by $p = -{A}/{\rho^B}$). The degeneracy between the two models at the level of the effective EoS parameter is qualitatively noteworthy. In this paper, we will discriminate quantitatively between the two models.


The adiabatic speed of sound squared for the Dieterici fluid (\ref{eq:Pde}) is given by
\beq
\label{soundd}
c_s^2:=\frac{\partial p}{\partial \rho}=A(B\rho+1) e^{B\rho}\,,
\eeq
which, for negative $A$ and $B$, is positive in the energy density range $\rho<-1/B$. We remark that outside this range, the model can still be stable under small-wavelength perturbations as explained in \cite{corr2}.

The cosmological relevance of Dieterici fluids can be better appreciated by formulating the model through a $k$-essence Lagrangian which thus is a function only of the kinetic energy of an underlying scalar field. An exact analytical result can be found in the non-relativistic regime, which can also open a path to the construction of the related Poincar\'e algebra \cite{poin} and allows for a transparent comparison with some other fluid approaches applied for the dark matter - dark energy unification in previous literature, namely the Generalized Chaplygin Gas and the Logotropic model. We recall that in the canonical formalism, the Lagrangian should be computed via the pressure \cite{pres1,pres2,pres3}:
\beq
\label{lanp}
{\mathcal L}=p\,.
\eeq
Introducing the scalar field $\theta$ which is canonically conjugate to $\rho$, i.e. their Poisson bracket reads as  $\{ \theta(x_i),\, \rho(x_j)   \}=\delta(x_i -x_j)$, the Lagrangian can be recast as
\beq\label{P-V}
{\mathcal L}(\rho,\dot\rho, t,x^i)=\dot\rho \theta -\frac{1}{2}\rho \partial_i \theta \partial^i \theta -V(\rho)\,.
\eeq
The potential for an irrotational fluid should be reconstructed by solving \cite{potpres}:
\beq
\label{pV}
p=\rho \frac{dV(\rho)}{d \rho} -V(\rho)\,,
\eeq
which, for a Dieterici fluid with $p=A\rho e^{B \rho}$, delivers
\beq
V(\rho)=-A\rho {\rm Ei}( - B \rho)\,,
\eeq
where the constant of integration has been fixed by imposing $V(0)=0$, and where ${\rm Ei}$ denotes the exponential integral. Thus:
\beq
\frac{dV(\rho)}{d \rho}=A(e^{B \rho} - {\rm Ei}(-B\rho) ).
\eeq
Next, the value of $\rho$ should be found by solving the Euler-Lagrange equation of motion
\beq\label{Euler}
\dot \theta= -\frac{1}{2} \partial_i \theta \partial^i \theta - \frac{dV(\rho)}{d \rho}\,.
\eeq
We will solve this equation of motion in the two limiting  regimes of low and high energies which correspond to dark energy and dark matter dominated epochs. We obtain\footnote{Here $\gamma=0.5772156649$ denotes the Euler constant, and $W(\chi)$ is the special Lambert function which is the inverse of $\chi e^{\chi}$ \cite{stegun}.}:
\begin{eqnarray}
\label{regimed1}
&& \frac{dV(\rho)}{d \rho}\Big|_{\rho <<1}  \simeq A(1+\gamma+\ln(-B\rho)) + 2AB \rho \quad \Rightarrow \quad \rho=\frac{1}{2B}W\left( -2 {\rm Exp} \left( -\frac{\dot \theta +\frac{1}{2}\partial_i \theta \partial^i \theta}{A}-(1+\gamma) \right) \right)\,, \\
\label{diffpot}
&& \frac{dV(\rho)}{d \rho}\Big|_{\rho >>1}  \simeq A e^{B\rho} \quad \Rightarrow \quad \rho=\frac{1}{B} \ln \left[-\frac{1}{A} \left( \dot \theta +\frac{1}{2}\partial_i \theta \partial^i \theta \right) \right]\,.
\end{eqnarray}
Implementing these results into (\ref{lanp}) by using the Dieterici equation of state we get:
\begin{eqnarray}\label{dmlan0}
&& {\mathcal L}\simeq \left[-W\left( -2 {\rm Exp} \left( -\frac{\dot \theta +\frac{1}{2}\partial_i \theta \partial^i \theta}{A}-(1+\gamma) \right) \right) {\rm Exp}\left(-\frac{1}{A} \left(\dot \theta +\frac{1}{2} \partial_i \theta \partial^i \theta \right)-(1+\gamma)\right)\right]^{\rm 1/2} \quad {\rm for} \quad {\rho<< 1}\,, \\
\label{dmlan}
&& {\mathcal L}\simeq \left( \dot \theta +\frac{1}{2}\partial_i \theta \partial^i \theta \right) \ln \left( -\frac{\dot \theta +\frac{1}{2}\partial_i \theta \partial^i \theta}{A}\right)   \quad {\rm for} \quad {\rho>> 1}\,,
\end{eqnarray}
where we have also taken into account that Lagrangians that differ by a multiplicative constant are physically equivalent (that is, they provide the same equations of motion). Our analysis shows that the parameter $B$, which quantifies the strength of the attractive effects between the fluid particles, does not affect the theory Lagrangian neither in the low energy limit nor in the high energy scenario. The physical meaning of the parameter $B$ has been provided in \cite{dou1,dou2}: we are considering a fluid formed by molecules that are interacting with each other.  We assume that this attractive interaction force is experienced only by molecules close to each other. Inside the gas, the mutual forces between the molecules balance each other, but
this would not be any longer the case for the molecules in the proximity of the boundary, in which case we will have a net force. Applying an argument by Boltzmann, it is shown that this net force does not influence the distribution of the velocities of the molecules (i.e., the kinetic energy), but their spatial distribution (i.e., the density). Thus, $B$ is a measure of the work that should be done for moving the molecules towards the boundary against this force field. Up to date, although a Lagrangian formulation has been proposed for the Generalized Chaplygin Gas, the physical meaning of its free parameters has not yet been clarified, unlike for the Dieterici model. Furthermore, from (\ref{diffpot}), we can understand that the functional form of the Dieterici equation of state leads to a shift in the derivative of the potential in the high energy scenario if compared to the logotropic framework \cite{luongo}. In this latter paper results have been derived by means of Taylor expansions around certain values of the free model parameters, while in our case, the series expansions have been carried out with respect to the values of the energy density.  Therefore, even though the Dieterici equation of state interpolates between dark matter and dark energy dominated epochs, as the Generalized Chaplygin Gas and the Logotropic model do (and they can be regarded as analogue proposals at the level of the effective fluid description), we can see that they deeply differ at the level of the underlying theory. In fact, in the former cases, the Lagrangian is monotonic with respect to the potential energy of the scalar field, while in the Dieterici case, non-monotonic behavior can be identified in (\ref{dmlan}) which represents the dark matter-dominated phase. 

Dubbing $X= \dot \theta +\frac{1}{2}\partial_i \theta \partial^i \theta$ the  kinetic energy, and expanding the   Lagrangian  (\ref{dmlan0})  about $X=0$, we obtain  at second order\footnote{Again here we use that Lagrangians which differ by multiplicative or additive constants are physically equivalent.}
\begin{equation}
{\mathcal L}\simeq X+C X^{2}+o\left(X^{3}\right)\,, \qquad
C=-\frac{s^3 +5 s^2 +6 s +4}{4\left(1+s\right)^{2} A\left(s+2\right)}\,, \qquad s=W\left(-2 \mathrm{e}^{-(1+\gamma)}\right)\,,
\end{equation}
where it should be emphasized that the deviations from  linearity depends on the parameter $A$ only. Furthermore, in the regime characterized by (\ref{regimed1}) the Dieterici equation of state parameter (\ref{w}) and the adiabatic speed of sound squared (\ref{soundd})  as  functions of the kinetic energy $X$ read as
\beq
w = A\left[-\frac{2 \mathrm{e}^{-\left(\frac{X}{A}+(1+\gamma)\right)}}{W\left(-2 \mathrm{e}^{-\left(\frac{X}{A}+(1+\gamma)\right)}\right)}\right]^{\frac{1}{2}}\,, \qquad
c_{s}^{2} = \frac{w}{2} \left[2+W\left(-2 \mathrm{e}^{-\frac{X}{A}-(1+\gamma)}\right)\right] \,,
\eeq
from which we can compute
\beq
\frac{ d c_s^2}{dX}= -\frac{y(4+y)}{4(1+y)}e^{\frac{y}{2}}    \,, \qquad y=W(-2 e^{-u})\,, \qquad u=\frac{X}{A}+1+\gamma\,.
\eeq
Therefore, for understanding whether the adiabatic speed of sound within the Dieterici fluid is a monotonic function of the underlying scalar field kinetic energy, we need to establish whether the function
\beq
\label{signcs}
\tilde s= \frac{W(z)[4+W(z)]}{1+W(z)}
\eeq
comes with a well-defined sign. We should note that $z=-2 e^{-u}<0$, and that for a negative argument, the above quantity switches its sign in correspondence of the likewise property of $1+W(z)$. Thus, $c_s^2$ is not a monotonic function of $X$. We should also note that the approximation we are using here is applicable as long as $z>-1/e$ as by definition of Lambert function, which means as long as $X<A[\ln(1/2e) +1+\gamma]$.

On the other hand, by  simple algebraic manipulations we get that the non-relativistic Lagrangian in (\ref{dmlan}) about $X=0$ is given by a softcore (e.g. logarithmic in the kinetic energy) correction to the linear regime:
\beq
{\mathcal L}\simeq X \left[1-\frac{ \ln (X)}{\ln |A|}\right] \,.
\eeq
In the regime characterized by (\ref{diffpot}) the equation of state parameter and sound speed squared respectively read as 
\begin{equation}
    \omega=-X\,, \qquad
c_{s}^{2}=-X \left[ \ln \left(-\frac{X}{A}\right)+1\right] \,.
\end{equation}
implying
\begin{equation}
    \frac{d c_s^2}{dX} =-\left[\ln \left(-\frac{X}{A}\right)+2 \right]\,.
\end{equation}
Therefore, in this regime, the adiabatic speed of sound is a monotonic function of the kinetic energy of the scalar field, which is a physical property shared by the Generalized Born-Infeld model \cite{avelino}. We should remark, however, that in the case of the Dieterici fluid, taking into account the discussion about (\ref{signcs}),  such a feature arises in the high energy regime only. This is an important difference between the physical foundations of the Generalized Chaplygin Gas and the Dieterici fluid, which was not self-evident by inspecting Fig. \ref{figeos}.

\subsection{Field theory formulation of the (Modified) Berthelot fluid}

We will now construct the underlying Lagrangian theory  for the (Modified) Berthelot fluid (\ref{EOS2}) in both non-relativistic  and relativistic regimes. We start by noticing that for this fluid the equation of state parameter and adiabatic speed of sound squared read respectively as
\begin{equation}
\label{wbert}
w=\frac{B}{1+A \rho}\,, \qquad
c_{s}^{2}=\frac{\partial p}{\partial \rho}=\frac{B}{\left(1+A \rho\right)^{2}}\,.
\end{equation}

\subsubsection{Non-relativistic regime}
In order to derive the non-relativistic Lagrangian formulation of which the (Modified) Berthelot fluid is the hydrodynamical realization, we follow again the classical formulation for an irrotational perfect fluid  by introducing the potential V($\rho$) of the scalar field $\theta$. We reconstruct the potential by substituting (\ref{EOS2}) into (\ref{pV})
\beq
V(\rho)=B \rho \ln  \left( {\frac {\rho}{A\rho+1}} \right),
\eeq
where we have fixed the constant of integration by imposing $\lim_{\rho \to 0^+}V(\rho)=0$. When studying the rotation curves of galaxy we will enlighten a relationship between this Lagrangian potential and the Newtonian gravitational potential in (\ref{potbert}). Next, by inserting
\beq
\label{dVnonr}
\frac{dV(\rho)}{d \rho}=B \ln \left(\frac{\rho}{A \rho+1}\right)+\frac{B}{A \rho+1}.
\eeq
into (\ref{Euler}) we can obtain the following relationship between fluid energy density and scalar field kinetic energy $X=\dot{\theta}+\frac{1}{2} \partial_{i} \theta \partial^{i}$:
\begin{equation}
\rho=-\frac{W\left[-A \operatorname{Exp}\left(-\frac{\dot{\theta}+\frac{1}{2} \partial_{i} \theta \partial^{i} \theta+B}{B}\right)\right]}{A\left(W\left[-A \operatorname{Exp}\left(-\frac{\dot{\theta}+\frac{1}{2} \partial_{i} \theta \partial^{i} \theta+B}{B}\right)\right]+1\right)}=-\frac{W\left[-A \operatorname{Exp}\left(-\frac{X+B}{B}\right)\right]}{A\left(W\left[-A \operatorname{Exp}\left(-\frac{X+B}{B}\right)\right]+1\right)}    \,.
\end{equation}
Now, plugging  the above result into (\ref{lanp}) we get for the (Modified) Berthelot  equation of state (\ref{EOS2}):
\begin{eqnarray}
&& {\mathcal L}= W\left[-A \cdot {\rm Exp} \left(-\frac{\dot \theta+\frac{1}{2}\partial_i \theta \partial^i \theta+B}{B}\right)\right] = W\left[-A \cdot {\rm Exp} \left(-\frac{X+B}{B}\right)\right],
\end{eqnarray}
where again we have taken into account that Lagrangians  differing by a multiplicative constant are physically equivalent because they provide the same Euler-Lagrange equations of motion. It should be noted that unlike in the Lagrangian formulation for the Dieterici fluid (\ref{dmlan0})-(\ref{dmlan}), in this case both the free model  parameters $A$ and $B$ enter the expression for the Lagrangian.  Expanding the  Lagrangian we have just constructed about   $X=0$  we get
\begin{equation}
{\mathcal L}\simeq X+C X^{2}+o\left(X^{3}\right)\,, \qquad
C=\frac{W\left(-A \mathrm{e}^{-1}\right)^{2}}{2\left(1+W\left(-A \mathrm{e}^{-1}\right)\right)^{4} B^{3}} \,,
\end{equation}
which shows that the dependence on both the free parameters survives also in this limit.
As from (\ref{wbert}) the equation of state parameter and speed of sound squared are related to the scalar field kinetic energy via 
\begin{equation}
w=  B [W(v)+1] \,, \qquad  c_s^2=B [W(v)+1]^2 \,, \qquad v=-A\mathrm{e}^{-\frac{B+X}{B}}\,.
\end{equation}
Therefore we can compute 
\beq
\frac{d c_s^2}{dX} =-2 W\left(v\right)
\eeq
which shows that $c_s^2$ is a monotonic function of $X$ because the Lambert function is a monotonic function of its argument and $v$ is a monotonic function of $X$. This property characterizes the fluid models belonging to the family of the Generalized Born-Infeld theories \cite{avelino}.

\subsubsection{Relativistic regime and relation to the Dirac-Born-Infeld theory}

\begin{figure}
    \centering
    \includegraphics [scale=0.45]{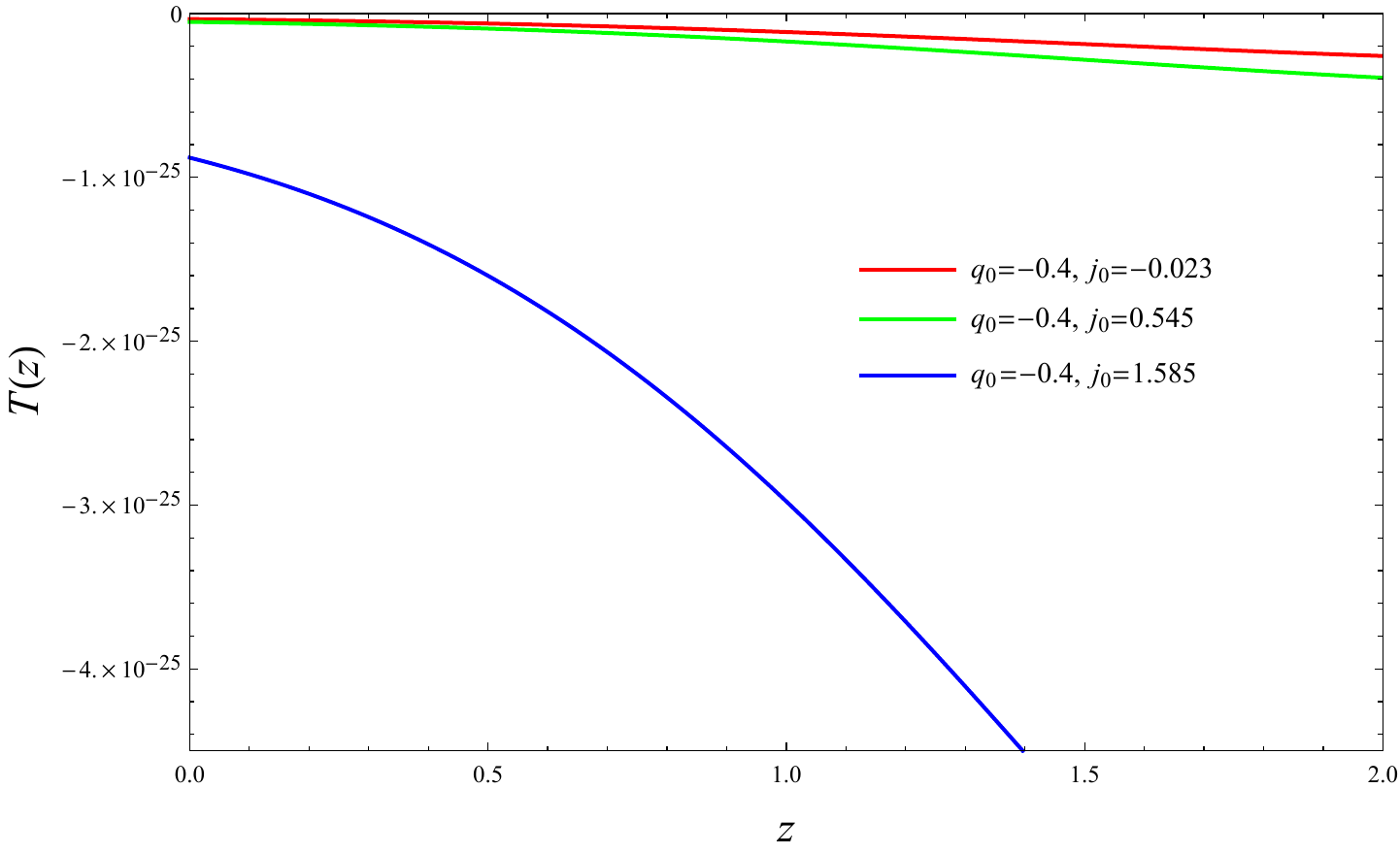}
    \caption{We display the redshift evolution of the tension of the brane $T(z)=1/f(z)$ we found in (\ref{tensioncosmo}) using the Hubble function $H= H(z)$  reconstructed from gaussian processes applied to cosmic chronometers data. We assume $H_0=64.9\pm4.2~kms^{-1}Mpc^{-1}$, $q_0=-0.4$, and $j_0=[-0.023,0.545,1.585]$ respectively. The tension of the brane is expressed in units of $kg/m^3$.}
    \label{fig:tension}
\end{figure}

The relativistic form  of the underlying  Lagrangian for the (Modified) Berthelot fluid can be constructed in the framework  of the Dirac-Born-Infeld (DBI) field theory proposed in string theory.  The general form of the DBI Lagrangian for a scalar field $\phi$ is \cite{tong}
\begin{equation}\label{DBIL}
\mathcal{L}_{D B I}=\frac{1}{f(\phi)}\left(\sqrt{1+f(\phi) \dot{\phi}^{2}}-1\right)-V(\phi) \,,
\end{equation}
where  V($\phi$) denotes the potential and the function $f(\phi)$ is interpreted as the inverse of the tension of the brane. In the Dirac-Born-Infeld theory the pressure  and energy density of the fluid are related to the scalar field via
\beq
\label{DBIbasics}
\rho=\frac{\gamma}{f(\phi)}+\left(V(\phi)-f(\phi)^{-1}\right) \,, \qquad
p =-\frac{1}{f(\phi) \gamma}-\left(V(\phi)-f(\phi)^{-1}\right) \,, \qquad \gamma = \frac{1}{\sqrt{1-f(\phi) \dot{\phi}^{2}}}\,,
\eeq
where the latter plays the role of a Lorentz-like contraction factor.
Should we assume a non-interacting scalar field with $V(\phi)=0$, we can easily invert the previous relations into
\begin{eqnarray}
\label{f(phi)}
f(\phi) &=& \frac{\rho-p}{p \rho} \,,  \\
\label{phidot}
\dot{\phi}^{2} &=& \frac{p(p+\rho)}{\rho} \,.
\end{eqnarray}
By using the chain rule for derivatives and the energy conservation equation, we can rewrite   (\ref{phidot}) as 
\begin{equation}\label{dphi/drho}
\frac{d \phi}{d \rho}= - \frac{1}{\rho} \cdot \sqrt{\frac{p}{3(\rho+p)}}\,,
\end{equation}
which, once (\ref{EOS2}) is implemented,  can be integrated into 
\begin{equation}
\label{phirho}
\phi(\rho)=2 \sqrt{\frac{B}{3 (B+1)}} \tanh ^{-1}\left(\sqrt{\frac{B+A \rho +1}{B+1}}\right)+C,
\end{equation}
where $C$ is a constant of integration which can be determined by imposing an initial condition $\phi\left(\rho=\rho_{0}\right)=\phi_{0}$. Next, inverting  this expression as $\rho=\rho(\phi)$ and plugging the result into (\ref{f(phi)})  we  obtain
\begin{equation}\label{f}
f(\phi)=\frac{A}{B(B+1)} \cdot\left[(B-1) \cosh^{2}\left(\frac{\phi -C}{2} \cdot \sqrt{\frac{3(B+1)}{B}}\right)+B+1\right]\,,
\end{equation}
which concludes the procedure for finding the Dirac-Born-Infeld formulation (\ref{DBIL}) for the (Modified) Berthelot fluid (\ref{EOS2}). By plugging (\ref{phirho}) into (\ref{f}) and using the Friedman equation, we obtain the inverse of the tension of the brane as a function of the Hubble function as:
\begin{equation}
   f(H)= \frac{A}{B}+\frac{8 \pi G(1-B)}{3 BH^2}\,,
\end{equation}
which can then be written in terms of the Hubble constant, of the deceleration parameter, and of the jerk parameter as
\begin{equation}
\label{tensioncosmo}
    f(j_{0},q_{0},H_{0},H)=\frac{8  \pi G \left[3 H^{2}\left(-j _{0}+q_{0}+2 q_{0}^{2}\right)+H _{0}^{2}\left(-4+3 j_{0}+3 q_{0}-4 q_{0}^{3}\right)\right]}{3 H^{2} H _{0}^{2}(1-2 q_{0})^{2}(1+q_{0})}\,,
\end{equation}
where a subscript $0$ indicates that the quantity is evaluated at the present time. We can note that the constant $C$ arising in (\ref{phirho}) does not affect the final result. We plot in Fig. \ref{fig:tension}  the redshift evolution of the tension of the brane where we assume the model-independent estimates of the Hubble constant \cite{H0_value} $H_{0}=64.9\pm4.2~kms^{-1}Mpc^{-1} \approx 2.10305 \times 10^{-18}s^{-1}$, and of the deceleration parameter \cite{q0_value} $q_0\approx-0.4$ from cosmic chronometers data; we implement the Hubble function $H=H(z)$ we have reconstructed by applying Gaussian processes to the cosmic chronometers data in our previous paper \cite{EPJC}. The three values for the jerk parameter have been taken from \cite{jerk0}, and they have been estimated by using Taylor, Padè, and Rational Chebyshev method, respectively.

\begin{figure}[!htb]
    \centering
    \includegraphics[scale=0.42]{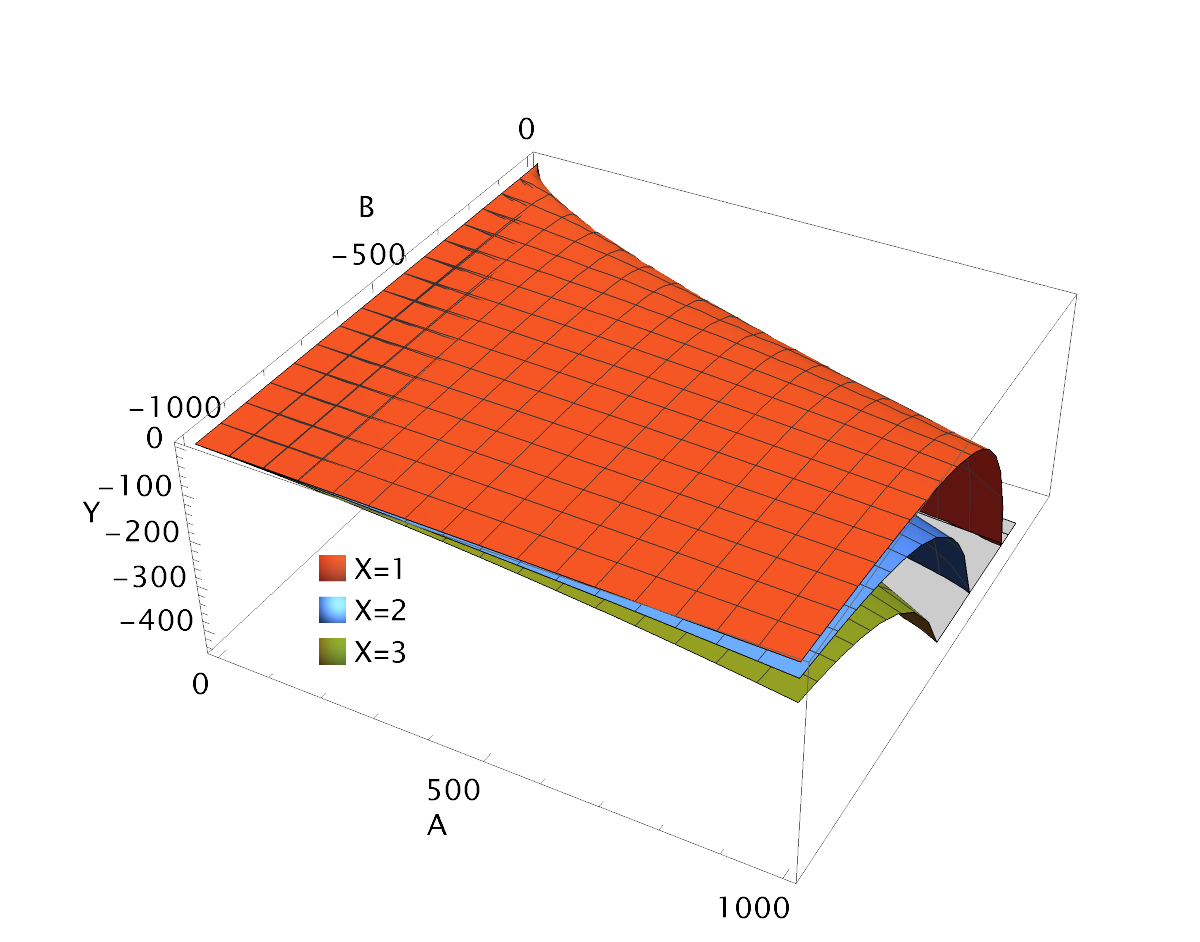}
    \caption{This figure shows the  behavior of $Y=4 A X-B^{2}-1+\sqrt{(B+1)^{2}-8 A X}(1-B)$} over A and B for three values of $X$, namely $X=[1,2,3]$.
    \label{ynumeric}
\end{figure}

For low kinetic energy $2 X= g^{\mu \nu} \phi_{, \mu} \phi_{, \nu}= \dot \phi^2 \approx 0$, the Lagrangian (\ref{DBIL}) is approximated by
\begin{equation}
\label{dbilow}
     \mathcal{L}_{DBI}= X-\frac{ f(\phi)}{2} X^{2}+o\left(X^{3}\right)\,,
\end{equation}
where the field and its kinetic energy (e.g. its time derivative) have been treated as independent quantities. This result enlightens the difference between the Dirac-Born-Infeld theory at the core of the (Modified) Berthelot formulation, and the Generalized Born-Infeld theory which instead is the framework behind the Generalized Chaplygin Gas \cite{avelino}. The latter equation of state reads as $p=-\frac{A}{\rho^\alpha}$ with $0<\alpha \leq 1$, and its corresponding Lagrangian has been reconstructed to be 
\begin{equation}\label{GBI}
\mathcal{L}_{G B I}=-A^{\frac{1}{1+\alpha}}\left[1-\left(2X\right)^{\frac{1+\alpha}{2 \alpha}}\right]^{\frac{\alpha}{1+\alpha}}\,,
\end{equation}
which belongs to the class of Generalized-Born-Infeld theory \cite{avelino}. Setting $\alpha=1$ we obtain both the original Chaplygin Gas formulation and the Born-Infeld theory, whose Lagrangian expanded for low kinetic energy provides
\beq
\mathcal{L}_{BI}= X+\frac{ X^2}{2} +o\left(X^{3}\right)\,,
\eeq
which is a particular subcase of (\ref{dbilow}). Therefore, from the reconstruction of the underlying Lagrangian we can conclude that the (Modified) Berthelot fluid, the Generalized Chaplygin Gas and the Anton-Schmidt fluid (see \cite{luongo}) all belong to the same family of the Chaplygin gas model as they extend the Born-Infeld theory although in different directions. Furthermore, combining (\ref{wbert}) and (\ref{phidot}) we can write the equation of state parameter and the adiabatic speed of sound of the (Modified) Berthelot fluid as functions of the scalar field kinetic energy as:
\begin{eqnarray}
w =  \frac{2(2AX-B)}{B-1+ \sqrt{(1+B)^2 -8AX}} \,, \qquad
c_{s}^{2} = \frac{4(2AX-B)^2}{(B-1+ \sqrt{(1+B)^2 -8AX})^2}\,,
\end{eqnarray}
where we have used
\beq
\rho(X)=\frac{B( B+1- \sqrt{(B+1)^2-8 A X})-4 A X}{2A(2AX-B)}\,.
\eeq
For establishing the monotonicity properties of the speed of sound, we note that $c_s^2(\rho)$ is monotonic increasing, and we can compute
\beq
\label{diffrhoX}
\frac{d \rho(X)}{dX}= \frac{B\cdot Y}{(2AX -B)^2 \sqrt{(B+1)^2-8 A X} } \,, \qquad Y= 4AX -B^2 -1 +\sqrt{(B+1)^2-8 A X} (1-B)\,, \qquad B<0\,,
\eeq
where, according to the numerical analysis reported in Fig. \ref{ynumeric}, $Y$ is negative.  Therefore, the sound speed squared $c_{s}^{2}=c_s^2(X)$  is  monotonically increasing.

By inverting (\ref{phirho}) into\footnote{We note that this formulation is valid as long as $B<-1$ because the cosmic fluid energy density needs to be positive. } 
\begin{equation}
    \rho(\phi)=-\frac{1+B}{A \cosh^2 \frac{(C-\phi) \sqrt{\frac{3(1+B)}{B}}}{2 }} \, ,
\end{equation}
we can as well write the equation of state parameter and the adiabatic speed of sound as functions of the underlying scalar field as: 
\begin{equation}
\label{w_cs}
w=\frac{B}{1-\frac{1+B}{ \cosh^2 \frac{(C-\phi) \sqrt{\frac{3(1+B)}{B}}}{2 }}}\,, \qquad
c_{s}^{2}=\frac{B}{\left(1-\frac{1+B}{ \cosh^2 \frac{(C-\phi) \sqrt{\frac{3(1+B)}{B}}}{2 }}\right)^2} \,.
\end{equation}
Thus, the adiabatic speed of sound $c_s^2$ of the (Modified) Berthelot fluid is monotonic with respect to the scalar field $\phi$ in the two ranges $\phi<C$ and $\phi>C$ separately being a composition of monotonic functions there. More specifically, it is increasing   in the former interval and decreasing in the latter.

\section{Testing the evolution of the Dieterici model at the background level}
\label{testing}

\subsection{Machine Learning approach and mock datasets}\label{sec:Meth}

When applying ML methods, we start by specifying the structure of the Neural Network, also referred to as \lq\lq the model", through a set of mathematical equations, and then we continue by running an optimisation algorithm which makes the Neural Network learn from the data it is provided with \cite{MLbook1,MLbook2}. In the specific Probabilistic Machine Learning, also known as Bayesian Machine Learning (BML), the learning process is repeated a large number of times for improving the ability of the Neural Network to perform its predictions, that is, to reconstruct the distribution of the posteriors. Furthermore, in this latter approach, the data are not taken from observations or experiments, but they are generated from the set of mathematical equations defining the model, that is, the model is  used also for organizing the generative process which precedes the learning process \cite{MLbook1,MLbook2}. For fixing the ideas, in this paper we will use the model Eqs.~(\ref{eq:F1})-(\ref{eq:Pde}) for generating the expansion rate data $H=H(z)$.

From the practical point of view, for carrying out the steps we have outlined, we exploit the probabilistic programming languages (PPLs) because they provide a complete framework for implementing BML and for computing the posterior distribution of the Neural Network parameters by applying the learning algorithm \cite{PPLs}. In particular, we  employ the python-based probabilistic programming package {\it PyMC3}  \cite{PyMC3} based on the deep learning library {\it Theano}\footnote{ \href{https://theano-pymc.readthedocs.io/en/latest/}{https://theano-pymc.readthedocs.io/en/latest/} }  which uses gradient-based Markov Chain Monte Carlo (MCMC) algorithms and Gaussian processes for generating the mock data.   
In our case,  Eqs.~(\ref{eq:F1})-(\ref{eq:Pde})  are employed in the generative process. Our analysis is based on $10$ chains with $3500$ generated mock datasets, where the observable is the Hubble rate as a function of the redshift $H=H(z)$.

Since these field equations can only be solved numerically, we  have decided to integrate them  with the help of the Runge–Kutta \lq\lq RK4" method \cite{RK} in which we have set the step-size $h = 0.001$. The RK4 method evolves the input for the initial values, such as ($ x_{n} $, $ y_{n} $),   into   ($x_{n+1}$, $ y_{n+1} $) by use of the following discretized system of equations:
\begin{eqnarray}
&&K_{1}=h \cdot f\left(x_{n}, y_{n}\right),\quad
K_{2}=h \cdot f\left(x_{n}+\frac{h}{2}, y_{n}+\frac{K_{1}}{2}\right),\quad
K_{3}=h \cdot f\left(x_{n}+\frac{h}{2}, y_{n}+\frac{K_{2}}{2}\right),\quad \nonumber\\
&&K_{4}=h \cdot f\left(x_{n}+h, y_{n}+K_{3}\right),\quad
y_{n+1}=y_{n}+\frac{K_{1}}{6}+\frac{K_{2}}{3}+\frac{K_{3}}{3}+\frac{K_{4}}{6}\,,
\end{eqnarray}
where $h\equiv \Delta z$, $z$ being the redshift in our case, and $ f(x,y) $ is the differential equation to solve, i.e. the Friedman equations (\ref{eq:F1}). Providing this algorithm with a set of initial conditions  we can obtain the values of the cosmological parameters, such as the Hubble function and the matter abundance, at a certain redshift.

In the next section, we will apply  BML to simulate the data $H=H(z)$ that will be used in the learning process and forecast the parameters of our cosmological model without the need for astrophysical data because the  BML algorithm is able to create a mock dataset by learning from the model itself (generative process). Then, cosmic chronometers and BAO data are used to validate our results.  We will carry out the generative step  for the model given by Eqs.~(\ref{eq:F1})-(\ref{eq:Pde})  over  two different redshift ranges: 

\begin{itemize}
	\item $z \in [0, 2.5]$ because this is the range in which the observational data for the cosmic history of the Universe are available, and they can be used for validating the results from BML. 
	
	\item $z \in [0, 5]$ because future explorations (like the one based on gamma-ray bursts  \cite{grb}) will cover this redshift range, and their observations may be used in subsequent papers to improve the validation of the BML results.
\end{itemize}

Once we are done with the mocking of the dataset and the simulation of the Hubble function $H=H(z)$, that we will need in the learning process, using  Gaussian processes, we use the available observational data about the cosmic history of the Universe to validate the BML's forecast on the Dieterici cosmological parameters. We take the observational data from \cite[Table~I]{zhang2016}. The values of the Hubble rate at different redshift reported in this dataset have been estimated   by implementing the   ages of passively evolving galaxies located at different redshift into the relation:
\begin{equation} \label{DvsH}
H(z)=-\frac{1}{(1+z)} \frac{d z}{d t} \approx-\frac{1}{(1+z)} \frac{\Delta z}{\Delta t}\,.
\end{equation}
We can note that this method does not require any pre-determined assumption about the properties of the cosmic fluid and that, unlike the measurements of the luminosity distance of supernovae sources, it is based on a {\it differential} (and not {\it integral}) procedure which preserves the information about the redshift evolution of the effective EoS parameter \cite{jimenez}.

\subsection{Forecasts for the cosmological parameters}\label{sec:results}

\begin{figure}[h!]
	\begin{center}$
		\begin{array}{cccc}
		\includegraphics[width=90 mm]{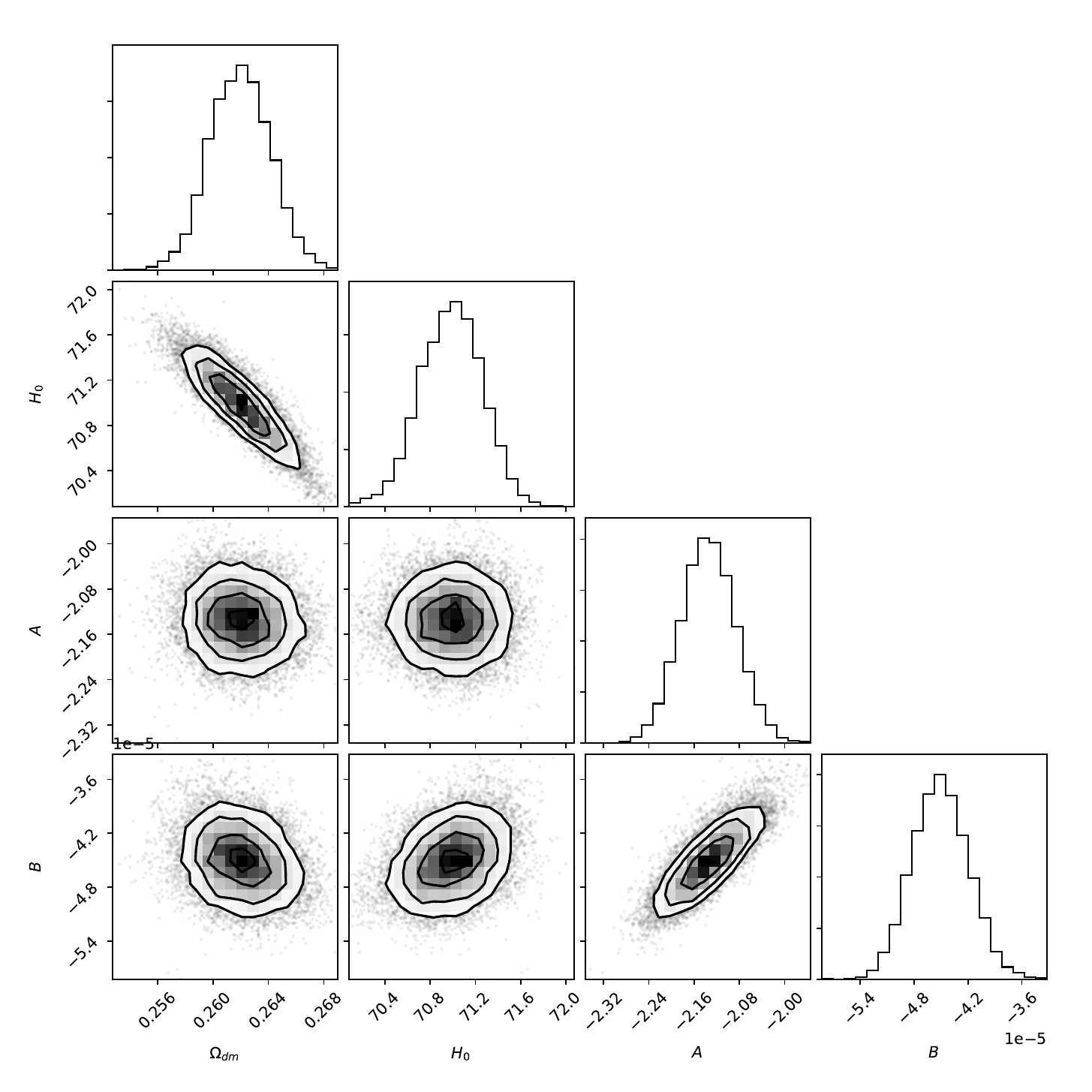}
		\includegraphics[width=90 mm]{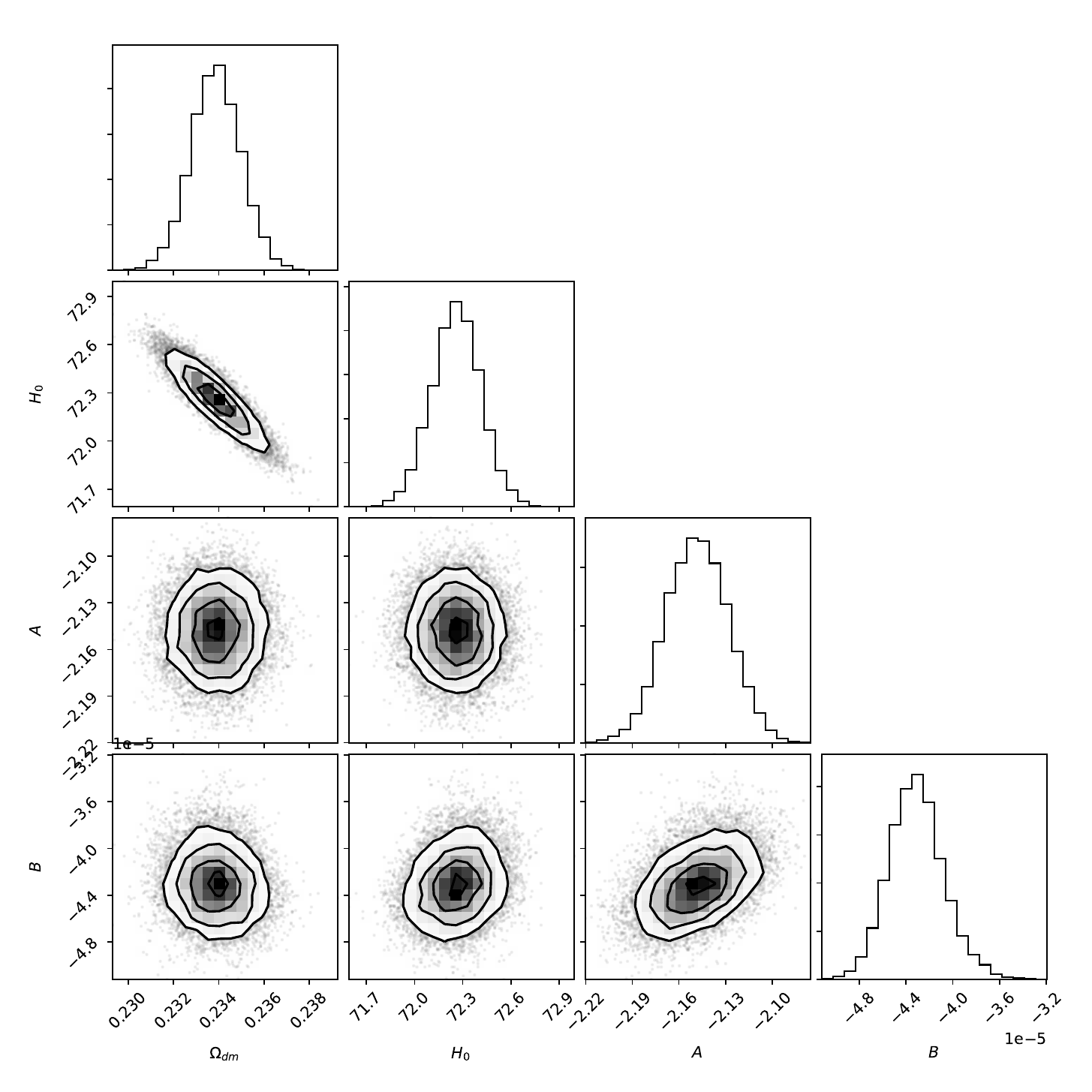}
		\end{array}$
	\end{center}
	\caption{ Forecasts on the Dieterici cosmological model given by Eqs.~(\ref{eq:F1})-(\ref{eq:Pde}) using mock data obtained from  the generative process applied to these two equations.  The left plot  shows the results when we consider the cosmic evolution for   $z \in [0,2.5]$, while in the right plot we have extended it to $z \in [0,5]$. The black contour lines show the $1\sigma$, $2\sigma$ and $2\sigma$ confidence regions, respectively.}
	\label{fig:Fig1_0}
\end{figure}

In this section, we exhibit the forecast bounds on the cosmological parameters entering the Dieterici model obtained applying the procedure described in Sect. \ref{sec:Meth}.
We have used the package {\it PymC3}  for  the generative process (which is based on  Eqs.~(\ref{eq:F1})-(\ref{eq:Pde})) imposing flat priors on the free cosmological parameters. We summarize our forecast for the cosmological parameters in  Table~\ref{tab:Table1}. It can be appreciated that  BML  imposes very tight constraints on the present-day value of dark matter abundance $\Omega_{d m 0}$, on the free parameters of the Dieterici EoS, and  on the Hubble constant  $H_{0}$.  Our results about the latter quantity suggest that a unified dark matter - dark energy paradigm pictured according to the Dieterici equation of state, in which the pressure decays exponentially with respect to the energy density, may mitigate the Hubble tension because values $H_0 \in ( 70, \, 72 ) $  km s$^{-1}$ Mpc$^{-1}$ are consistent with WMAP and South Pole Telescope estimates based on Cosmic Microwave Background datasets \cite[p.11]{eleonora}.  Then, we exhibit in Fig.~\ref{fig:Fig1_0} (left panel) the contour plots of various combinations of the cosmological parameters when the data have been generated in the interval $z \in[0,2.5]$.
BML imposes tight constraints on the cosmological parameters when we generate the data in the redshift range $z \in[0,5]$ as well. In this latter case, we obtain a smaller central value for $\Omega_{d m 0}$.   In Fig.~\ref{fig:Fig1_0} (right panel), we present the contour plots of various combinations of cosmological quantities for the analysis in this latter redshift range. Also in this case, we have assumed flat priors for the cosmological parameters.

\begin{table}[ht]
	\centering
	\begin{tabular}{|c|c|c|c|c|} 
		\hline
		$z$  & $\Omega_{dm0}$ & $H_{0}$ & $A$ & $B$ \\
		\hline
	&&&&\\[-1em]
		 $z\in[0,2.5]$ & $0.262 \pm 0.002$ &$70.992^{+0.278}_{-0.264}$ km/s/Mpc & $-2.135 \pm 0.048$ &  $-4.506 \times 10^{-5} \pm 0.314 \times 10^{-5}$  \\
		[3pt]
		\hline
		&&&&\\[-1em]
		$z\in[0,5]$  & $0.234 \pm 0.001$ & $72.261^{+0.154}_{-0.146}$ km/s/Mpc & $-2.148 \pm 0.019$ & $-4.317 \times 10^{-5} \pm 0.235 \times 10^{-5} $ \\
		[3pt]
		\hline
	\end{tabular}
	\caption{The best fit values for the cosmological model given in Eqs.~(\ref{eq:F1})-(\ref{eq:Pde}) with the corresponding $1\sigma$ errors  when BML has generated the data in the two different intervals $z \in [0,2.5]$ and  $z \in [0,5]$, respectively. We have searched the best fit values in the  ranges $\Omega_{dm0} \in [0.2,0.5]$, $H_{0} \in [62.0,85.0]$ $km/s/Mpc$, $A \in [-5.0,0.0]$ and $B \in [-0.5,0.5]$ imposing flat priors to the cosmological parameters.}
	\label{tab:Table1}
\end{table} 

\begin{figure}[h!]
	
	\includegraphics[width=75 mm]{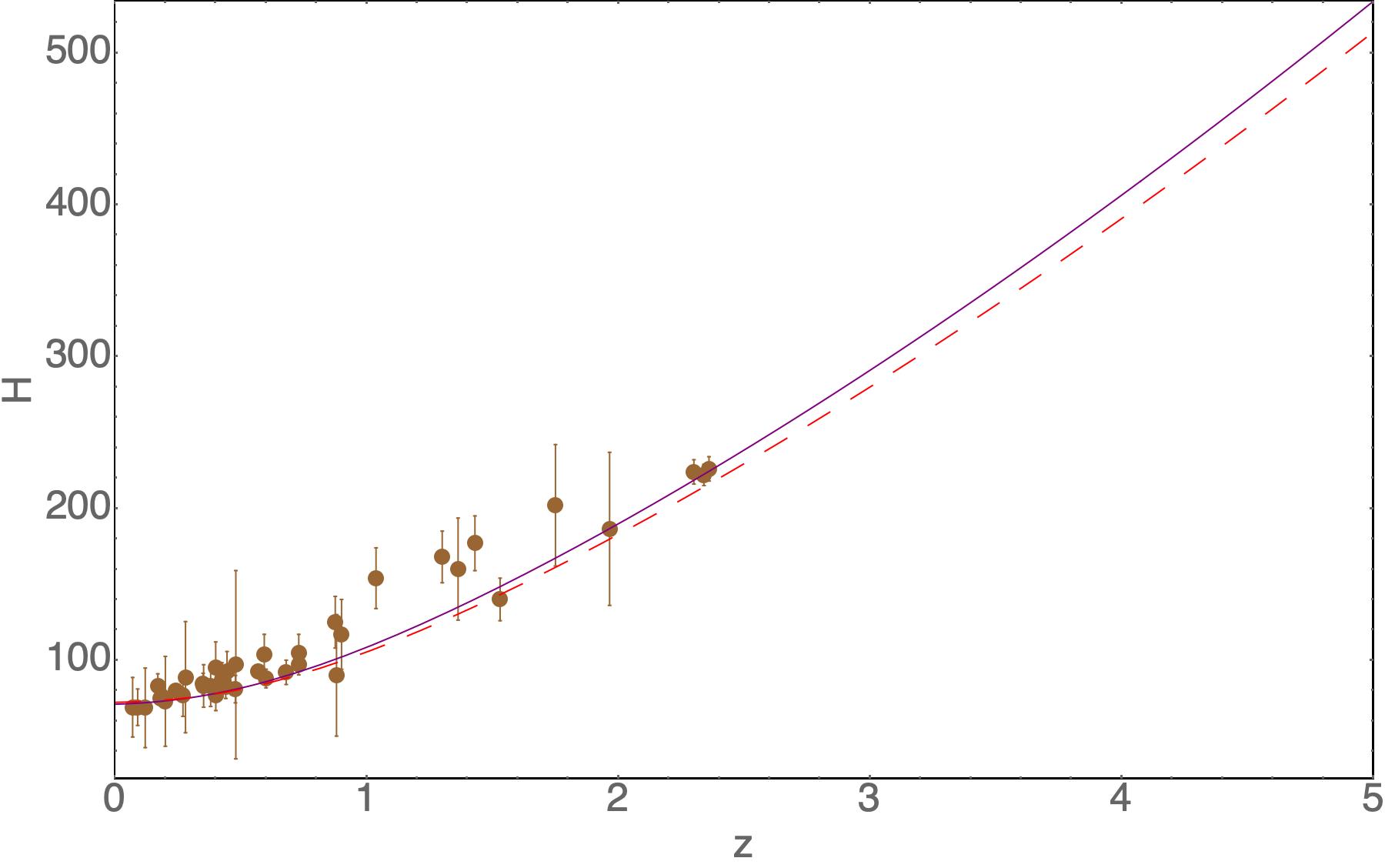}
	
	\caption{The BML prediction for the redshift evolution of the Hubble function  $H=H(z)$ compared to the cosmic chronometers and BAO data (red dots) from \cite[Table~I]{zhang2016}. The purple curve has been obtained when the data are generated in $z \in [0,2.5]$, while the dashed red curve takes into account the data generated in $z \in [0,5]$.  The free parameters affecting the two curves have been fixed as reported in Table~\ref{tab:Table1}.}
	\label{fig:Fig1_1}
\end{figure}

In order to validate the outcome of our analysis, we compare in Fig.~ \ref{fig:Fig1_1} the redshift evolution of the Hubble function from BML  with the available combined observational data of cosmic chronometers and BAO.  In the plot, the purple curve refers to the case in which we have generated the data for $z \in [0, 2.5]$, while the dashed red curve holds when $z \in [0, 5]$. The dots are the 40 data points from astrophysical observations taken from  \cite[Table~I]{zhang2016}. 
We can note that the BML result for the redshift evolution of the  Hubble function is in good agreement with the observational data at low redshift, but some tension could arise at high redshift. Indeed, the two estimates of $H_0$ reported in Table~\ref{tab:Table1} are not consistent with each other at 1$\sigma$ level.

\begin{figure}[h!]
	\begin{center}$
		\begin{array}{cccc}
		\includegraphics[width=75 mm]{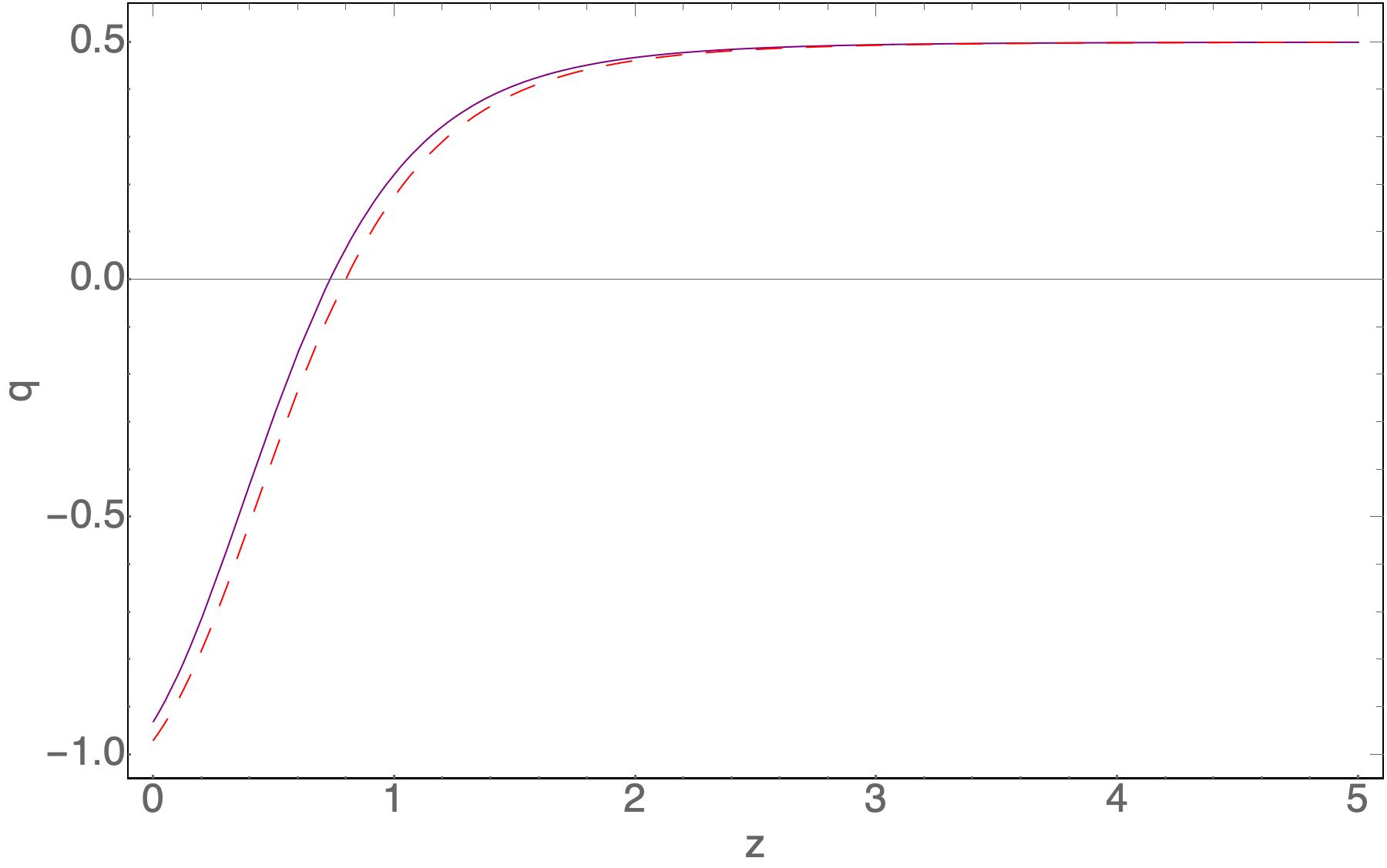}&&
		\includegraphics[width=75 mm]{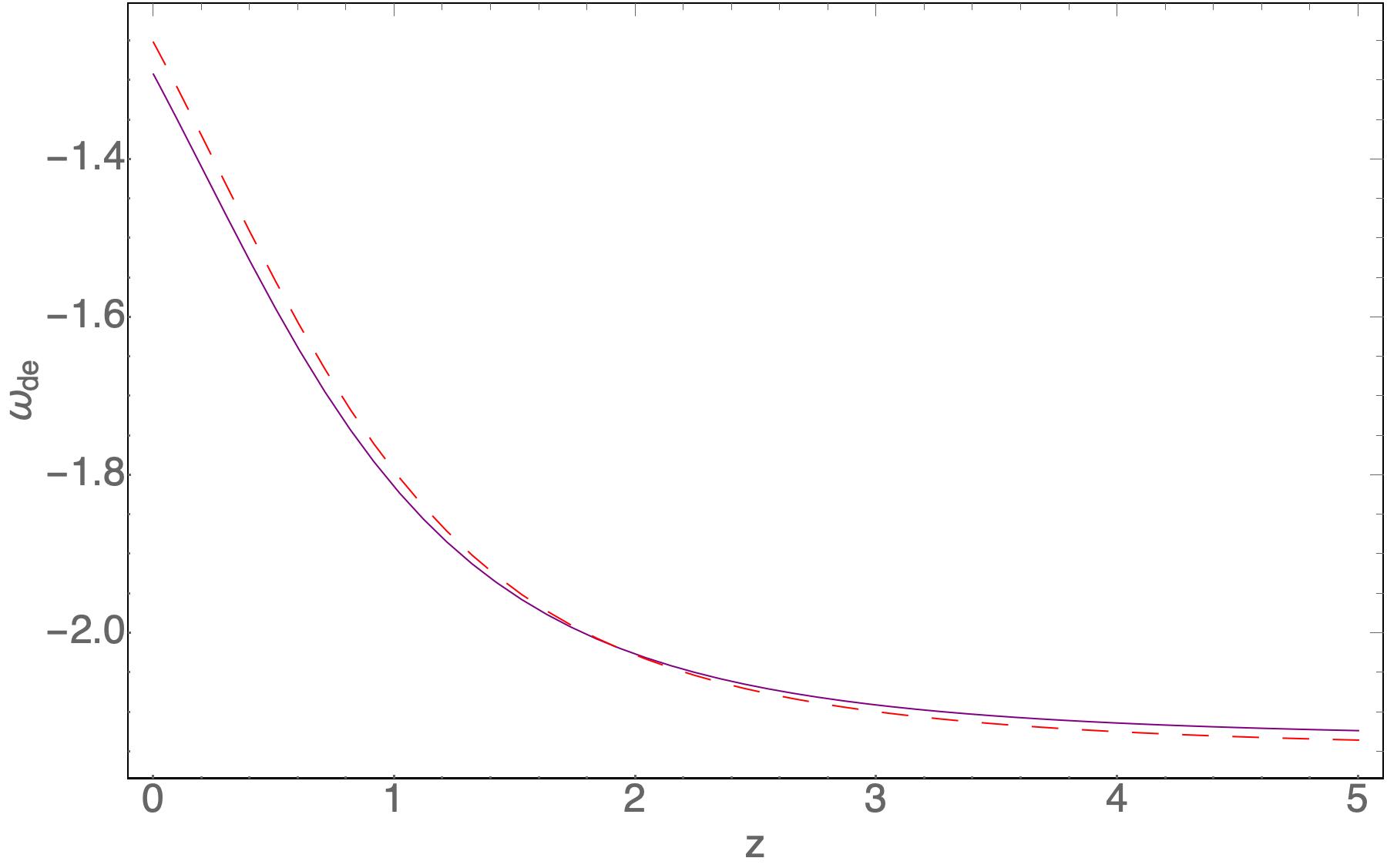} \\
		\end{array}$
	\end{center}
	\caption{ BML predictions for the redshift evolution of the EoS parameter $w$ (right panel) and deceleration parameter $q$ (left panel). The same conventions on the types of the curves as in Fig.~\ref{fig:Fig1_1} apply. }
	\label{fig:Fig1_2}
\end{figure}

In Fig.~\ref{fig:Fig1_2}, we  display the redshift evolution of the deceleration parameter $q$ (left panel) and of the effective EoS parameter of the Dieterici fluid $w$ (right panel), for the best fit values of the model parameters exhibited in Table~\ref{tab:Table1}. We use the same convention for the purple and dashed red curves as in Fig.~\ref{fig:Fig1_1}. We can see that the BML analysis predicts a  transition between a decelerating and an accelerating phase of the Universe. More quantitatively, we obtain that the transition to a negative deceleration parameter, e.g. to the present epoch of accelerated expansion of the Universe,  occurs at a lower value of the redshift, e.g. more recently, when our analysis relies on the data in $z \in [0, \, 2.5]$. In other words, the Dieterici model exhibits a correlation between a more recent transition and a lower value of the Hubble constant $H_0$ and an higher value of the present day dark matter abundance $\Omega_{dm0}$.  Furthermore, from the plot in the right panel, one can understand that the Dieterici fluid behaves as a phantom fluid at the present epoch because  $w<-1$.

\section{On some cosmic model degeneracy, and its breaking  }

\subsection{Cosmic opacity and model degeneracy}
\label{opacity}

The dark energy abundance can be estimated by analyzing the luminosity distance data of type Ia supernovae; historically, this was the first astrophysical observation that had called for the introduction of such a mysterious fluid. Therefore, ultimately it is necessary to reconstruct the flux of photons arriving from those sources:  light rays travel along geodesics experiencing the spacetime curvature, which is assumed constant in the usual Friedman-Lema\^itre-Robertson-Walker model. It has been claimed that non-constant curvature effects along the photons' path may mimic some observational properties attributed to dark energy \cite{inhom1,inhom2,inhom3,inhom4,inhom5,inhom6,inhom7}. On the other hand, scattering and absorption effects with the interstellar medium can affect the propagation of light rays as well. This means that photons do not travel through a perfectly transparent region whose opacity should be considered as a redshift dependent function $\eta=\eta(z)$. According to this  framework, the dark energy abundance $\Omega$ and its equation of state parameter $w(z)$ should be found by comparing astrophysical data against the corrected luminosity distance function, which, for for a flat universe, reads
\beq
D_L = (1+z) \eta(z)  \int_0^z \frac{dz'}{\sqrt{(1-\Omega) (1+z')^3+ \Omega {\rm Exp}\left[3 \int_0^{z'}\frac{1+w(\bar z)] d \bar z}{1+ \bar z} \right]}}\,,
\eeq
where the idealized case of a perfectly transparent universe is recovered for $\eta(z) \equiv 1$. A plethora of different possibilities for the opacity function\footnote{Some other authors prefer to work with the function $\tau=\tau(z)$ such that $\eta(z)=e^{\tau(z)/2}$.  } $\eta(z)$ has been proposed in the literature for implementing information about many different phenomena which may affect the travel of photons as scattering, absorption, and gravitational lensing effects experienced by the light rays when travelling inside some dust clouds or nearby massive astrophysical objects.  For example, in \cite{Lee} the applicability of some proposals of the cosmic opacity as
$\tau(z)=2 \epsilon z$,  $\tau(z)=(1+z)^{2 \epsilon}-1$,  $\eta(z)=\eta_{0}$, 
$ \eta(z)=\eta_{0}+\eta_{1} z $, $\eta(z)=\eta_{0}+\eta_{1} z /(1+z)$, 
$\eta(z)=\eta_{0}+\eta_{1} z /(1+z)^{2}$, 
$\eta(z)=\eta_{0}-\eta_{1} \ln (1+z)$, 
$\eta(z)=\eta_{0} /(1+z)$, 
$\eta(z)=\eta_{0} /(1+z) \exp [z /(1+z)]$, 
$(1+z)^{\eta_{0}}$, where $\epsilon$, $\eta_0$ and $\eta_1$ are  free model parameters, have been tested in light of relevant astrophysical datasets as Union2, SDSS, 6dFGS, WiggleZ, BOSS, BAO, etc... . Already in the past decade there were investigations on the redshift evolution of the cosmic opacity \cite{eta1,eta2,eta3,eta4} similarly to what done for the dark energy equation of state parameter  \cite{w1,w2,w3,w4}; recently these studies  have acquired a novel relevance in light of the Hubble tension issue \cite{PyMC3_p2}.  

\begin{figure}
    \centering
    \includegraphics[scale=0.38]{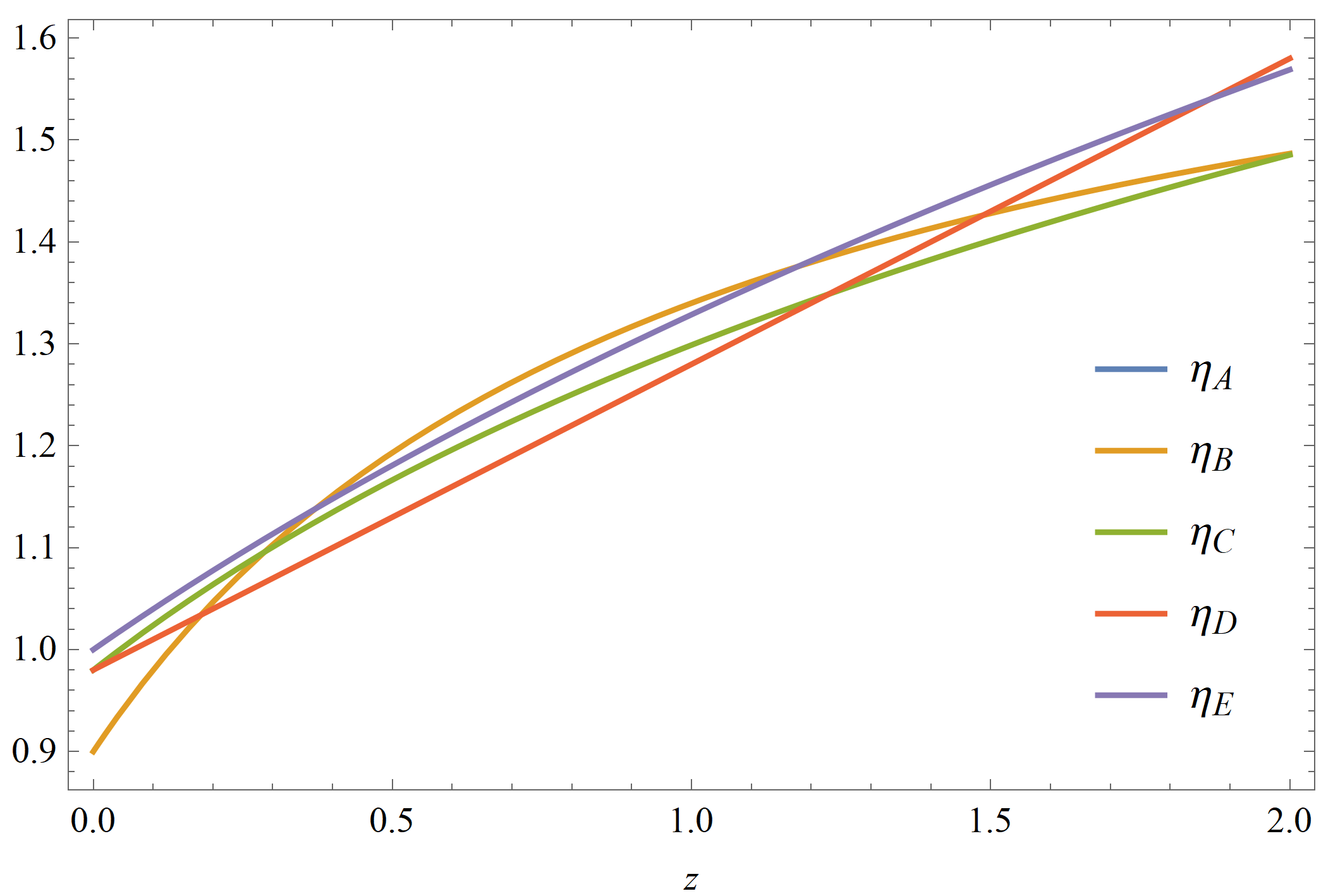}
    \caption{The cosmic opacity function $\eta_A(z)$ (\ref{etaA}),  which would lead to a degeneracy between the luminosity distances of the Milne and Einstein-De Sitter universes, is compared with some other cosmic opacity functions proposed in other literature, namely $\eta_B=\eta_0+\eta_1 z/(1+z)$, $\eta_C=\eta_0-\eta_1 \ln(1+z)$ with $[\eta_0 \approx 0.90,\, \eta_1 \approx 0.88]$ and $[\eta_0 \approx 0.98,\, \eta_1 \approx -0.46]$ respectively, $\eta_D=\eta_0 +\eta_1 z$ with $[\eta_0 \approx 0.98,\, \eta_1 \approx 0.30]$, and $\eta_E=(1+z)^{\eta_0}$ with $\eta_0 \approx 0.41$. Qualitative similarities are noteworthy.  }
    \label{fig:degeneracy}
\end{figure}

However, the following problem arises:
it is conceivable that completely different pairs ($\tau(z)$, $w(z)$) will perform statistically equivalently with the cosmic opacity function somehow masking the properties of the dark energy fluid. Let us explain what we mean by considering the following simple example of known analytical luminosity distances; modulo some technicalities, the same reasoning can be extended to the dark energy fluid models. The luminosity distances in the Milne and Einstein-De Sitter universes are respectively given by \cite{{Hogg}}
\begin{equation}
d_{L1}(z)=z\left(1+\frac{z}{2}\right)\,, \qquad
d_{L2}(z)=2(1+z-\sqrt{1+z})\,.
\end{equation}
By multiplying $d_{L2}(z)$ by the cosmic opacity function
\beq
\label{etaA}
\eta_A (z)= \frac{z(2+z)}{4(1+z-\sqrt{1+z})}
\eeq
we will get the luminosity distance $d_{L1}(z)$. Next, we should note that $\eta_A(z)$ can be well-approximated by some cosmic opacity functions which had already arisen in previous literature suggesting that it is not a completely  ad-hoc result. See Fig. \ref{fig:degeneracy}, in which we compare graphically $\eta_A(z)$, with $\eta_B=\eta_0+\eta_1 z/(1+z)$,
$\eta_C= \eta_0-\eta_1 \ln(1+z)$ with $[\eta_0 \approx 0.90,\, \eta_1 \approx 0.88]$ and $[\eta_0 \approx 0.98,\, \eta_1 \approx -0.46]$ respectively, $\eta_D=\eta_0 +\eta_1 z$ with $[\eta_0 \approx 0.98,\, \eta_1 \approx 0.30]$, and $\eta_E=(1+z)^{\eta_0}$ with $\eta_0 \approx 0.41$.

Therefore, we must find some specific signatures for breaking the degeneracy between cosmological models based on different pairs cosmic opacity - dark energy equation of state. In the next section, we will tackle this problem by applying perturbation techniques because they are sensitive to the dark energy equation of state only. Consequently, our reasoning is analogue to what explored in \cite{Chakraborty} in which the degeneracy at a background level between $\Lambda$CDM and a specific $f(R)$ modified gravity model was broken by tracking the evolution of the density perturbations;  this way of thinking has already helped also in differentiating between dark interactions \cite{differentiating}. Likewise, it has been possible as well to discriminate between some $f(T)$ gravitational theories and their background equivalent dark energy fluid models  by looking at growth of perturbations \cite{referee}. 

\subsection{Breaking the model degeneracy through perturbation analysis}
\label{perturbation}

The equation governing the evolution of the density perturbation is given by \cite[Chapert 14]{WE}
\begin{equation}\label{perturb}
\ddot{\delta}+A(t) H \dot{\delta}+\left(  B(t)+\frac{c_{s}^{2} n^{2}}{H^{2} a^{2}}\right) H^{2} \delta=0,
\end{equation}
where $\delta$ is the density fluctuation,  $n$ is the wave-number of the Fourier mode of the perturbation, and 
\beq
 A(t)=2-3 \omega-3\left(\omega-c_{s}^{2}\right)\,, \qquad
 B(t)=-\frac{3}{2}\left[(1-\omega)(1+3\omega)+6\left(\omega-c_{s}^{2}\right)\right] \Omega+\frac{12 k}{H^{2} a^{2}}\left(\omega-c_{s}^{2}\right) \,,
\eeq
where $\omega=p/\rho$, $c_s^2=\frac{\partial p}{\partial \rho}$ and $k$ is the spatial curvature of the Friedman universe.

Perturbation analysis for a universe filled by the Chaplygin gas  has revealed that a phase of growing perturbations occurs at early times, which is followed by a decreasing oscillatory phase at late times asymptotically going to zero \cite{P.chap}. On the other hand,  for Logotropic and Anton-Schmidt fluids, considering the best-fit parameters with respect to Pantheon supernovae data, observational Hubble data, and $\sigma_8$ data points, it has been pointed out in \cite{anton} that matter perturbations decrease in time and remain bound, except for the model in which the Grüneisen parameter is fixed at  priori in which case the perturbations are diverging at early times. For these latter models, there is no oscillatory behavior. 

If we consider a flat  single-fluid universe  (for which $\Omega=1$) and introduce the energy contrast defined as $\xi=\delta(t)/\rho(t)$, by expanding the time derivatives and eliminating the ones of the energy density via $\dot\rho=-\sqrt{3\rho(t)} (\rho(t)+p(t))$, we can express (\ref{perturb}) in the long-wavelength limit (i.e. $n/(Ha) \ll 1$) as 
\begin{equation}\label{density}
\ddot{\xi}+\sqrt{\frac{\rho(t) }{3}} (3 c_s^2-12 \omega -4) \dot{\xi}+   \left(3 c_s^2+9 \omega ^2+6 \omega +2\right) \rho(t)\xi=0 \,,
\end{equation}
where we have eliminated the Hubble function by using the Friedman equation $H(t)=\sqrt{\rho(t)/3}$. We will now discuss in some detail the evolution of the density contrast in universes filled by the (Modified) Berthelot and the Dieterici fluids separately focusing on their weakly nonlinear regimes for which analytical solutions in terms of special functions can be found. We will as well focus on the effects beyond the background level.

\begin{figure}[h!]
	\begin{center}$
		\begin{array}{cccc}
		\includegraphics[width=85 mm]{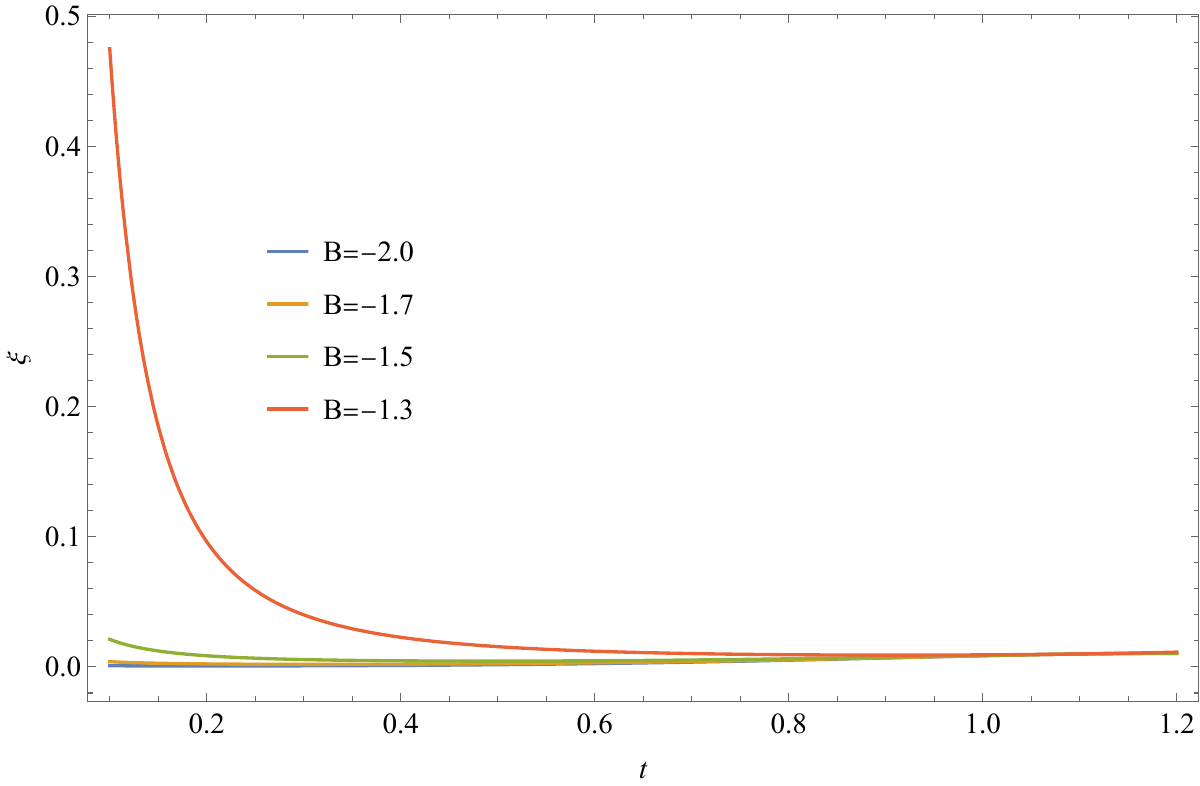}&&
		\includegraphics[width=85 mm]{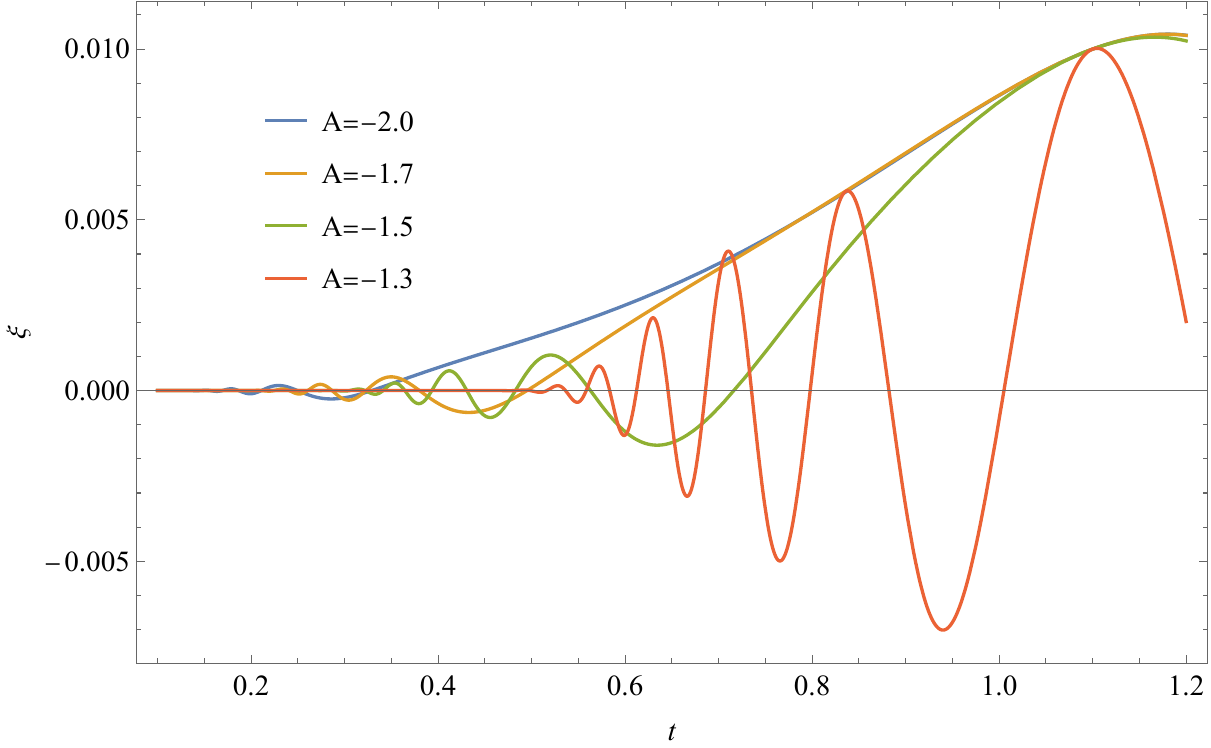}\\
		(a)    && (b)
		\end{array}$
	\end{center}
	\caption{The time evolution of the density contrast $\xi=\delta/\rho$ for a single fluid flat universe filled with (Modified) Berthelot fluid (panel a) and Dieterici fluid (panel b) for four values of B=[-2.0,-1.7,-1.5,-1.3] and A=[-2.0,-1.7,-1.5,-1.3], respectively for the two models. In panel (a), we  have integrated backwards in time eqs. (\ref{density2})-(\ref{terms}) with initial conditions $\xi(1.10)=0.01$,  $\dot{\xi}(1.10)=0.01$ and fixing $A=0.1$. In panel (b),  we  have integrated backwards in time eqs.(\ref{density2})-(\ref{terms2}) with the same initial conditions as panel (a) but fixing $B=-0.1$.}
	\label{fig:xi.plot}
\end{figure}

Beginning with the (Modified) Berthelot fluid, we approximate the solution to its background energy conservation equation as  $\rho(t) \sim \frac{4}{3 t^{2}(B+1)^{2}}$. Implementing it into (\ref{density}) and keeping only the leading order terms in $A$ we obtain
\begin{eqnarray}
\label{density2}
&& A(t)\ddot{\xi}+B(t)  \dot{\xi}+ C(t)\xi=0 \,,\qquad
 A(t)=\frac{t^{2}(B+1)^{2} \left(8 A+3(B+1)^{2} t^{2}\right)}{3}\,, \\
\label{terms}
&&  B(t)=-\frac{2 t(B+1) [3(1+B)^2 (9B+4)t^2 +16A(3B+2)]}{9}\,, \qquad  C(t)=\frac{4(3 B+2)\left(8 A+3(3 B+1)(B+1)^{2} t^{2}\right)}{9}.
\end{eqnarray}
This equation can be solved analytically as
\begin{equation}
\xi(t)=C_1 t^{\frac{15B+11+\beta}{6 \left(B+1\right)}} \cdot h_1(t) +C_2 t^{\frac{15 B + 11-\beta}{6 \left(B+1\right)}}  \cdot  h_2 (t)
\end{equation}
where $C_1$ and $C_2$ are constants of integration, $\beta=\sqrt{225 B^{2}+186 B+25}$, and
$ h_1$ and $h_2$ are the following hypergeometric functions:
\begin{eqnarray}
h_1(t) &=& _2F_1\left(\left[\frac{\beta-9 B-5}{12( B+1)},\, \frac{5+\beta-3B}{12( B+1)}\right],\,\left[\frac{6( B+1)+\beta}{6( B+1)}\right];\, -\frac{3 \left(B+1\right)^{2} t^{2}}{8 A}\right)\,, \\
h_2(t) &=& _2F_1\left(\left[\frac{5-3B-\beta}{12( B+1)}, \,-\frac{5+3B+9\beta}{12 (B+1)}\right],\,\left[\frac{(6 B+1)-\beta}{6( B+1)}\right]; \, -\frac{3 \left(B+1\right)^{2} t^{2}}{8 A}\right)\,.
\end{eqnarray}
For an intuitive visual representation of the evolution of the density contrast in a universe filled with a (Modified) Berthelot fluid with small non-linearities, we integrate backwards in time (\ref{density2})-(\ref{terms}) with initial conditions $\xi(1.10)=0.01$,  $\dot{\xi}(1.10)=0.01$, and with $A=0.1$ and various choices for $B$; the result is presented in Fig.~\ref{fig:xi.plot} (panel a). We obtain that for all values of the parameter $B$ we have investigated, the density contrast $\xi$ is monotonically decreasing in time in the interval considered.

For the case of a universe filled with a Dieterici fluid, inserting again the approximated result for the energy conservation law at the background level $\rho(t) \sim \frac{4}{3 t^{2}(A+1)^{2}}$ into (\ref{density}), and approximating  $B\approx 0$, we arrive at an equation for the evolution of density contrast analogous to (\ref{density2}) but with the following time-dependent coefficients:
\begin{eqnarray}\label{terms2}
&& A(t)=9 t^6 (1 + A)^6 \,, \qquad
B(t)=-2 (A +1) t \left(8 A  B ^2+3 (9 A +4) (A +1)^4 t^4+24 A  (A +1)^2 B  t^2\right)\,, \\
&& C(t)=4 \left(8A (12A+5) B ^2+3 (1+3A)(2+3A) (A +1)^4 t^4+24A (3A+2) (A +1)^2 B  t^2\right)\,. \nonumber
\end{eqnarray}
In this case, the general solution for the evolution of the density contrast  is\footnote{ We should note that the solution is valid for $A \neq -1$ only because the differential equation with coefficients (\ref{terms2}) loses meaning in that case due to the vanishing of two of them. The same behavior arises also in the context of the (Modified) Berthelot fluid (\ref{density2})-(\ref{terms}) for $B=-1$. }:
\begin{eqnarray}
&&\xi(t)=\left[C_{3}t^{\frac{12 A+8}{3 (A+1)}} \cdot {\rm HeunB}\left(-a,b,c,d;\frac{2 \sqrt{A} B}{3(A+1)^{5 / 2} t^{2}}\right) + C_{4}t^{\frac{3 A+1}{A+1}}\cdot {\rm HeunB} \left(a,b,c,d;\frac{2 \sqrt{A} B}{3(A+1)^{5 / 2} t^{2}}\right)\right] \cdot {\rm Exp}\left( -\frac{4\left[6t^2(A+1)^2+ B\right] A B}{9(A+1)^{5} t^4}\right)\,,\nonumber\\
&& a=\frac{3 A+5}{6( A+1)}, \qquad b= 4\sqrt{\frac{A}{A+1}}, \qquad c=\frac{37(3 A+1)}{6( A+1)},\qquad d=-\frac{2\left(9 A+7\right) \sqrt{A}}{3(A+1)^{3 / 2}}\,,
\end{eqnarray}
where  $C_3$ and $C_4$ are constants of integration and HeunB denotes the Heun's Biconfluent special function.
For an intuitive visualization, we display the evolution  of the density contrast in a universe filled with a Dieterici fluid with initial conditions $\xi(1.10)=0.01$,  $\dot{\xi}(1.10)=0.01$, setting $B=-0.1$ and assuming various values of $A$ in Fig.~\ref{fig:xi.plot} (panel b). We obtain that the density contrast $\xi$ exhibits an increasing and oscillatory behavior in time for  the values of $A$ and the time interval we have considered. We can therefore appreciate important qualitative differences between panels (a) and (b), e.g. between the evolution of the density contrast for the (Modified) Berthelot and Dieterici fluids. The instabilities identified in the latter scenario might be related to the corresponding evolution of the speed of sound, which governs their velocity of propagation, and which contains an exponential factor $\sim e^{\rho} \sim e^{1/t^2}$. This factor amplifies the (negative) value of the adiabatic speed of sound in the time interval we have considered, and can actually serve as an hint for identifying this suitable time interval necessary for breaking the degeneracy.


\section{$k$-essence and free Dirac-Born-Infeld as dark matter candidates only}
\label{rotation}

In this section, we will assess whether the two fluid models we have previously introduced, namely the Dieterici and (Modified) Berthelot equations of state, can account or not for flat galactic rotation curves. We are here assuming that these models can be adopted for a description of the dark matter content of the universe, and this section should be considered as in parallel to the ones in which those models have instead been invoked as candidates for the dark energy budget. Indeed, there has been a recent interest in quantifying the effects of a non-zero adiabatic speed of sound of dark matter, as arising in our models, on galactic scales \cite{malafarina}. In particular, we will show that the free Dirac-Born-Infeld model cannot account for flat rotation curves and a specific condition on the potential for achieving this goal is proposed. The gravitational potential withing the galaxy will also be written in a closed form with respect to the energy density allowing us to discuss a possible physical reason on the different applicability of the two models.

\subsection{Galaxies Rotation Curve in the Dieterici and (Modified) Berthelot frameworks}

Following \cite[Eqs.(5)-(6)-(7)]{malafarina}, we describe the dark matter density profile far away from the center of the galaxy adopting the newtonian equations of hydrodynamic equilibrium, which read as:
\begin{eqnarray}
\label{eqmass}
\frac{d M(r)}{dr} &=& 4 \pi r^2 \rho (r) \,, \\
\label{condP}
\frac{d P(r)}{dr} &=& -\rho(r) \frac{M(r)}{r^2}   \,, \\
\label{condpot}
\frac{d \Phi(r)}{dr} &=& \frac{M(r)}{r^2} \,,
\end{eqnarray}
where $M(r)$ is the mass of the galaxy contained in a sphere of radius $r$, and $\Phi(r)$ is the gravitationl potential. Then, it can be shown that the rotation velocity is \cite{malafarina}
\beq
\label{velocity}
v = \sqrt{\frac{M(r)}{r}} \,.
\eeq
Eq.(\ref{condP}) can be rewritten as
\beq
\label{condP1}
c_s^2 \frac{d \rho(r)}{dr} = -\rho(r) \frac{M(r)}{r^2}\,,
\eeq
which has been used for assessing the effects of a non-zero speed of sound of dark matter in the galactic rotation curves \cite{malafarina,kuantay}, and which for the case of the Dieterici fluid becomes explicitly
\beq
\label{diett}
 \frac{d \rho(r)}{dr} = - \frac{ \rho(r)\, M(r) \, e^{-B \rho(r)}}{A r^2 (1+B \rho(r))}\,.
\eeq
Differentiating both sides of (\ref{eqmass}) with respect to $r$, plugging in (\ref{diett}), and finally eliminating the energy density via $\rho(r) = \frac{d M(r)}{ 4 \pi r^2 dr}$ we obtain the differential equation 
\beq
\label{massdieterici}
4 \frac{d M(r)}{dr} {\rm Exp} \left[ - \frac{B \, d M(r)}{4 \pi r^2 dr}  \right]  \pi r M(r) +A \left( r \frac{d^2 M(r)}{dr^2} -2 \frac{d M(r)}{dr}\right)\left( 4 \pi r^2 +B \frac{d M(r)}{dr} \right)=0\,,
\eeq
which governs the radial behavior of the mass of the galaxy. Next, we assume a power-law profile with $M(r) \sim r^\alpha$ into the previous condition arriving at
\beq
4 \pi r^\alpha {\rm Exp} \left[ - \frac{ B \alpha r^{\alpha -3}}{4 \pi}  \right] +A(\alpha -3) (B \alpha r^{\alpha -2} + 4\pi r)=0\,.
\eeq
For obtaining a flat rotation curve at spatial infinity by using (\ref{velocity}), we need to require $\alpha=1$, with the previous condition now reading
\beq
4 \pi r e^{- \frac{B}{4 \pi r^2}} -2A \left( \frac{B}{r} + 4\pi r \right)=0\,,
\eeq
whose asymptotic behavior is 
\beq
4 \pi (1-2A) r +O \left( \frac{1}{r} \right) =0\,,
\eeq
showing that a Dieterici fluid model can account for asymptotically flat galactic rotation curves with (in natural units)\footnote{Here, our purpose is to provide a critical overview of the applicability of these fluid models as dark energy {\it or}   dark matter separately, {\it or} a unification of the two. Thus, the finding of a positive value of $A$ is not in contradiction with the previously presented analysis on the cosmic history, but it should be interpreted as suggesting that a $k$-essence fluid of this type can also be used as a modeling of dark matter on galactic scales in which case dark energy  should be accounted for with some other mechanism. We apply the  same way of thinking also to the (Modified) Berthelot fluid.}  $A=1/2$. 

On the other hand, the explicit realization of (\ref{condP}) for a (Modified) Berthelot fluid is
\beq
\frac{d \rho(r)}{dr}=- \frac{\rho(r) M(r) (1+A \rho(r))^2}{r^2 B}\,,
\eeq
and Eq.(\ref{massdieterici}) should now be replaced by
\beq
16 B r^6 \pi^2 \frac{d^2 M(r)}{dr^2} + 16 \pi^2 r^4 \left( M(r) - 2B r \right) \frac{d M(r)}{dr} +8A \pi r^2 M(r) \left( \frac{d M(r)}{dr} \right)^2 +A^2 M(r) \left( \frac{d M(r)}{dr} \right)^3\,.
\eeq
Assuming again a power-law profile $M(r) \sim r^\alpha$ with $\alpha=1$, we obtain 
\beq
32 \left( \frac{1}{2} -B \right) \pi^2 r^4 + 8A \pi r^2 +A^2=0\,,
\eeq
which shows that a (Modified) Berthelot fluid cannot account for flat galaxy rotation curves. Hence, $k$-essence models and free Dirac-Born-Infeld models exhibit completely different applicability as dark matter candidates.

Finally, from (\ref{condpot}),(\ref{condP1}) we can write a differential equation for the gravitational potential with respect to the fluid density
\beq
\frac{d \Phi(\rho)}{d \rho} = - \frac{c_s^2}{\rho}\,,
\eeq
which, recalling (\ref{soundd}), (\ref{wbert}),  can be solved as
\begin{eqnarray}
\Phi(\rho) &=&  A [  {\rm Ei}(1,-B\rho)  - e^{B \rho}] + {\mathcal C}_1 \,, \\
\Phi(\rho) &=& B \left[ \ln \frac{1+A \rho}{\rho} - \frac{1}{1+A\rho} \right] + {\mathcal C}_2 \,,
\end{eqnarray}
for the Dieterici and the (Modified) Berthelot fluids respectively, where ${\mathcal C}_1$ and ${\mathcal C}_2$ are some constants of integration. We further notice that
\begin{eqnarray}
\frac{d \Phi}{d c_s^2} &\equiv&  \frac{d \Phi(\rho)}{d \rho} \cdot \frac{1}{\frac{d c_s^2}{d \rho}}    \, = \, - \frac{B \rho +1}{B \rho (2 +B \rho)}   \,, \\
\frac{d \Phi}{d c_s^2} &\equiv&  \frac{d \Phi(\rho)}{d \rho} \cdot \frac{1}{\frac{d c_s^2}{d \rho}}    \, = \, \frac{1+A\rho}{2 A \rho}>0 \,,
\end{eqnarray}
for the two equations of state respectively. While in the latter case $\Phi=\Phi(c_s^2)$ is an increasing monotonic function, this is not in the former case: this can constitute the physical reason behind the different clustering properties of the two fluids.  As an interesting result, recalling (\ref{dVnonr}),  we can  also relate the Newtonian potential within the galaxy to the Lagrangian scalar field potential in the non-relativistic regime for the (Modified) Berthelot fluid:
\beq
\label{potbert}
\Phi(r)= - \frac{dV}{d \rho}\,.
\eeq

\subsection{Galaxies Rotation Curves in the DBI framework}

Exploiting that the (Modified) Berthelot fluid is the hydrodynamical realization of the free Dirac-Born-Infeld model, we could claim that the latter is challenged by  astrophysical requirements of flat galactic rotation curves.  Therefore, we would like now to investigate whether these observations can be used for reconstructing some information on the potential of the scalar field.
Requiring an almost vanishing pressure $p=0$ in (\ref{DBIbasics}), we obtain the condition
\beq
V(\phi)= \frac{2 \gamma X}{1+\gamma} \,,
\eeq
or in other words
\beq
\label{solx}
X= V(\phi) \left( 1- \frac{V(\phi) f(\phi)}{2}  \right) \,.
\eeq
This requirement sets a constraint between the kinetic energy $X$ and the scalar field itself $\phi$, and therefore for  the cosmic energy density  $\rho(X,\phi)\,\, \to \,\, \rho(\phi)$. We interpret this result as imposing a relationship such that in the dark matter dominated era the time evolution of the potential is such that $V(t)=\frac{2 \gamma X(t)}{1+\gamma}$; see for example  \cite[Eq.(20)]{bertacca} and the $f=f(\phi(X))$ formalism which follows from imposing a certain condition on the pressure.    Explicitly we obtain
\beq
\label{solrr}
\rho(\phi)=   \frac{[2 - f(\phi)V(\phi)] V(\phi)}{1-f(\phi)V(\phi)} \,.
\eeq
The adiabatic speed of sound in (\ref{DBIbasics}), once  (\ref{solx}) is implemented, reads as
\beq
\label{cp}
 c_s ^2 = 1-f(\phi) V(\phi) \,.
\eeq
We note that
\beq
\dot X= \dot \phi \ddot \phi \qquad  \Rightarrow \qquad \ddot \phi = \frac{\dot X}{\dot \phi} = \frac{\frac{d X}{d\phi} \cdot \dot \phi}{\dot \phi} = \frac{d X}{d\phi}\,,
\eeq
where the relationship (\ref{solx}) is understood. Requiring a dark-matter-like behavior also for the speed of sound (\ref{cp}), e.g. $c_s^2 \approx 0$, we further write $V(\phi)= \frac{b}{f(\phi)}$ with $b$ some constant whose value in natural units is close to 1. Therefore, we obtain
\beq
c_s \approx \sqrt{1-b} \,, \qquad H = \sqrt{\frac{\rho(\phi)}{3}} \approx \sqrt{\frac{(2 - b)b}{ 3 (1-b) f(\phi)}} \,, \qquad \frac{d X}{d \phi} \approx \frac{f'(\phi)}{f^2 (\phi)} \left( \frac{b}{2} - 1 \right)      \,.
\eeq
Under our assumptions, the equation of motion of a Dirac-Born-Infeld scalar field (see e.g. \cite{DBI_inf1})
\beq
\ddot \phi + 3H c_s^2 \dot \phi + c_s^3 V'(\phi) + \frac{f'(\phi)}{2f(\phi)} \left( 1-\frac{2 c_s^2}{1 +c_s} \right) \dot \phi^2 =0 \,,
\eeq 
where a prime denotes a derivative with respect to $\phi$, becomes
\beq
\sqrt{1-b}[4b^2 - 5b - 2 - (2+b) \sqrt{1-b}] f'(\phi)    + 2 \sqrt{3} b(b-2)(1-\sqrt{1-b})    f(\phi) \approx 0\,,
\eeq
which comes with the solution
\beq
\label{fff}
f(\phi) = C e^{-D \phi} \,, \qquad D \equiv \frac{ 2 \sqrt{3} b(b-2)(1-\sqrt{1-b})}{\sqrt{1-b}[4b^2 - 5b - 2 - (2+b) \sqrt{1-b}]} \,,
\eeq
where $C$ is an arbitrary constant of integration.
The equation for hydrostatic equilibrium can be rewritten in terms of the scalar field, rather than of the energy density, as
\beq
\frac{d \phi}{d r} = \frac{1}{\frac{d \rho}{d \phi}} \cdot \frac{d \rho}{dr} \approx \frac{f(\phi) M(r)}{(1-b) f'(\phi) r^2} = \frac{ M(r)}{(b-1) D r^2}\,.
\eeq
Taking a derivative with respect to the radial coordinate $r$ and using (\ref{eqmass}) with $\rho \approx \frac{(2-b) b}{(1-b) f(\phi(r))}$ we obtain the second order differential equation governing the radial behavior of the scalar field: 
\beq
\frac{ d^2 \phi}{d r^2} + \frac{2}{r} \frac{d \phi}{d r} + \frac{4 \pi b (2-b)}{(1-b)^2 D\,C} e^{D \phi} =0\,.
\eeq
By introducing the auxiliary variable $ \phi (r) = \ln Y(r) $, the condition for hydrostatic equilibrium for the scalar field becomes
\beq
Y(r) \frac{d^2 Y(r)}{dr^2}  - \left( \frac{dY(r)}{dr} \right)^2 + \frac{2Y(r)}{r} \frac{d Y(r)}{dr} +  \frac{4 \pi b (2-b)}{(1-b)^2 D\,C} Y^{D+2}(r)=0 \,.
\eeq
Substituting a power-law solution $Y(r) \sim r^j$ we get $ j +  \frac{4 \pi b (2-b)}{(1-b)^2 D\,C} r^{Dj + 2} \sim 0$ which, by dimensional analysis, delivers $j=-2/D$, and therefore $\phi(r) \sim \ln r^{-2/D} $. Then, recalling (\ref{fff}), which provides $f(\phi(r)) \sim C r^2$,  and (\ref{solrr}) we obtain the radial profile of the energy density as
\beq
\rho(r) \sim \frac{(2-b) b }{(1-b) C r^2} \,.
\eeq
It can be appreciated that among the various possible dark matter profile, e.g. the ISO profile, exponential sphere, Burkert profile, NFW profile, Moore profile, Einasto profile, ours is well approximated at large distances from the center by the former \cite{ISOprofile}:
\beq
\rho_{\rm ISO} (r) = \frac{\rho_0}{ 1 + \left( \frac{r}{r_0}\right)^2} \, \to \, \frac{\rho_0 r_0}{r^2} \,, \qquad \rho_0 r_0 = \frac{(2-b) b }{(1-b) C }\,.
\eeq
Then, from  (\ref{eqmass}) and (\ref{velocity}) we get the rotation velocity of the galactic curve to be 
\beq
\label{vDBI}
v \approx \sqrt{\frac{4 \pi (2-b) b }{(1-b) C }} \,.
\eeq
Deepening the exploration of our Dirac-Born-Infeld model as {\it dark matter candidate only}, we discuss here the formation of cosmic structures, and then use this result jointly with the previous analysis for estimating the dark matter speed of sound within the galaxy. 
We begin by writing the  comoving Jeans length in the matter dominated era as \cite[Sect.IV]{logofereira}
\beq
\label{jeans}
\lambda^c_J = c_s (1+z) \left( \frac{\pi}{\rho} \right)^{1/2}\,.
\eeq
In the matter-dominated era, in which   $ p \approx 0$ and $c_s^2 \approx 0$ hold, we have $\rho(X) \approx \frac{2 X}{1-b} = \frac{\dot \phi^2}{1-b}$. Therefore, from the Friedman equation and (\ref{DvsH}) we arrive at
\beq
\frac{dz}{dt}= - \frac{(1+z) \dot \phi }{\sqrt{3(1-b)}}\,, \qquad \Rightarrow \qquad \phi(z)= - \sqrt{3(1-b)} \, \ln (1+z)\,.
\eeq
Still using the conditions stemming from $ p \approx 0$ and $c_s^2 \approx 0$, and eqs.(\ref{solrr}), (\ref{fff})  we get from (\ref{jeans})
\beq
\lambda^c_J  \approx (1-b) \sqrt{\frac{ \pi C}{(2-b)b}} (1+z_c)^{1-\frac{D \sqrt{3(1-b)}}{2}} \,,
\eeq
where $z_c \approx 1$ \cite[Sect.IV]{logofereira}. Isolating for $C$, taking into account (\ref{vDBI}), we can finally write for $b\approx 1$
\beq
c_s \approx \frac{\lambda^c_J v}{2 \pi (1+z_c)}\,.
\eeq
Re-introducing the Hubble constant for working in International System Units, and considering now $\lambda^c_J \approx R \approx 8 h^{-1}$,  $h \approx 0.73$ and $v \approx 130$ km/s, we get $c_s \approx 113$ km/s in good agreement with the result for the adiabatic speed of sound of dark matter in the halo  \cite[Table III]{malafarina}.

\section{Conclusion}\label{sec:conc}

Although the Dirac-Born-Infeld theory has been originally proposed for   taming the divergence of the electron self-energy as computed in classical
electrodynamics \cite{ED1,ED2}, it has received novel attention in recent decades in cosmological applications. One or multiple \cite{DBI_inf2} scalar field(s), possibly non-minimally coupled to other kinetic terms \cite{DBI_inf3}, to radiation \cite{DBI_inf4}, or to perfect fluids \cite{DBI3}, arising in the DBI framework may govern the inflationary epoch  in the slow-roll regime \cite{string2}. In this paper, we have proposed a Dirac-Born-Infeld interpretation for the fluid model (\ref{EOS2}) by assuming a vanishing potential: since the applicability of that fluid model was already put forward in \cite{capo},  our analysis may complement that of \cite{DBI2} in which specific DBI potentials were shown to give raise to fluid models challenged by observational datasets. Our result for the relation between the adiabatic speed of sound and DBI scalar field (\ref{w_cs}) for the (Modified) Berthelot fluid may be used in future projects for further assessing this specific realization of the cosmic matter content in light of the primordial black hole mass function invoking the sound speed resonance mechanism following \cite{DBI_inf1}, or tracking the amplification of the curvature perturbations as in \cite{tronconi}.  Models belonging to the DBI family may also be applied for describing other thermal epochs of the universe  \cite{DBI4,DBI5} either accounting for the dark energy abundance solely \cite{DBI6},  also specifically reproducing the cosmological constant term \cite{DBI8},   or for its unified form with dark matter \cite{DBI7}. This class of models provides various opportunities for realizing an accelerated phase of expansion of the universe  \cite{DBI9} even possibly taming the coincidence problem \cite{DBI13}. Being interested in this paper in an application in late-time cosmology, we have computed the relationship between the tension of the DBI brane generating the (Modified) Berthelot fluid model and the cosmographic parameters in (\ref{tensioncosmo}) and presented its redshift evolution in Fig. \ref{fig:tension}. The most important take-at-home messages  at the core of this manuscript, which can play a role as well in future applications, are: {\it (i)} the (Modified) Berthelot fluid is the hydrodynamic realization of the free Dirac-Born-Infeld theory; {\it (ii)}  the free Dirac-Born-Infeld theory cannot account for flat galactic rotation curves; {\it (iii)}  an exponential potential inversely proportional to the brane tension can fix this problem at a satisfactory level as shown by the predicted value of the adiabatic speed of sound of dark matter in the halo.

Furthermore, in this paper, we have also shown that another cosmological model proposed in \cite{dieterici1} as a dark matter - dark energy unification, e.g. the Dieterici one, does not challenge the Generalized Chaplygin Gas in the description of the cosmic history of the Universe. Even though the qualitative characteristics of the two models are similar, that is, they can both interpolate between decelerating and accelerating phases of the Universe, our quantitative analysis has shown that the cosmic chronometers and BAO data do not favor the former. In fact, the values of the Hubble constant estimated when we have generated the data in different redshift ranges are inconsistent with each other at 1$\sigma$ level (see   Table~\ref{tab:Table1}). We have also found that the Dieterici fluid would behave as a phantom fluid leading to a Big Rip singularity at which the predictability of the theory breaks down (should a Big Rip singularity occur, null and timelike geodesics would be incomplete \cite{bigrip1}). We would like to mention that however a phantom behaviour has been obtained also when the dark energy EOS is parametrized according to the Chevallier-Linder-Polarski (CPL) model \cite{Cap}: whether a formal connection exists between these two frameworks can be explored in a future project.
Our analysis has been performed using BML techniques already exploited by one of us in investigations about the opacity of the Universe, and the viscosity of the dark energy fluid \cite{PyMC3_p1,PyMC3_p2,PyMC3_p3}, and it belongs to the much wider research line which has been trying to apply techniques from data science to string theory \cite{string}. In fact, as summarized above,  in this paper, we have also put the fluid models we have considered in light of their scalar field interpretations, and possible degeneracy between physically different frameworks due to possible cosmic opacity terms have been broken by tracking the development of matter inhomogeneities.



\section*{Acknowledgement}
DG is a member of the GNFM working group of Italian INDAM.  MK acknowledges support from the program Unidad de Excelencia Maria de Maeztu CEX2020-001058-M and by the Juan de la Cierva-incorporacion grant (IJC2020-042690-I).

\section*{Data Availability Statement}
The data used in this manuscript have been taken from  \cite{H0_value,q0_value,jerk0}.

\end{document}